\documentclass[final,5p,times,twocolumn]{elsarticle}

\usepackage{amssymb}
\usepackage{amsmath,subcaption}
\usepackage{dcolumn,diagbox}%
\usepackage{natbib}

\begin{document}

\begin{frontmatter}



\title{Fast Super Robust Nonadiabatic Geometric Quantum Computation}


\author{Yi-fu Zhang}
\author{Lei Ma\corref{cor1}}
\ead{lma@phy.ecnu.edu.cn}
\affiliation{organization={School of Physics and Electronic Science},
            addressline={East China Normal University}, 
            city={Shanghai},
            postcode={200241}, 
            country={China}}
\cortext[cor1]{Corresponding author}
\begin{abstract}
Nonadiabatic geometric quantum computation (NGQC) provides a means to perform fast and robust quantum gates. To enhance the robustness of NGQC against control errors, numerous solutions have been proposed by predecessors. However, these solutions typically result in extended operation times for quantum gates. In order to maintain the robustness of quantum gates to control errors while shortening operation times to minimize the effects of decoherence, we introduce Fast Super Robust NGQC(FSR-NGQC). This approach achieves faster speeds when operating small-angle rotation gates. Through numerical calculations, we have demonstrated the performance of our scheme in a decoherence environment. The results show that our scheme achieves higher fidelity, thus enabling fast and robust geometric quantum computing.
\end{abstract}




\end{frontmatter}


\section{Introduction}
To efficiently execute quantum algorithms, it is important to reduce the error rate of quantum gates and increase coherence time. Whether they are NISQ algorithms\cite{preskillQuantumComputingNISQ2018,bhartiNoisyIntermediatescaleQuantum2022} that do not use error correction codes, such as VQE\cite{higgottVariationalQuantumComputation2019,tillyVariationalQuantumEigensolver2022} and QAOA\cite{zhouQuantumApproximateOptimization2020}, or future fault-tolerant quantum computation\cite{eganFaulttolerantControlErrorcorrected2021} schemes, there is a high demand for the fidelity of quantum gates. In recent years, scholars have made significant efforts toward this goal, achieved remarkable results in finding suitable physical systems\cite{ciracQuantumComputationsCold1995,clarkeSuperconductingQuantumBits2008,golterOptomechanicalQuantumControl2016,levineHighFidelityControlEntanglement2018} and optimizing quantum gate implementation schemes\cite{yanTunableCouplingScheme2018,niuUniversalQuantumControl2019}.

Among the numerous theoretical designs for optimizing quantum gate implementations, the application of quantum geometric phases\cite{zhangGeometricHolonomicQuantum2023} has proposed a new pathway to construct quantum gates with high robustness. Geometric phases are divided into two categories: Abelian geometric phases\cite{berryQuantalPhaseFactors1984,ekertGeometricQuantumComputation2000} and the more complex non-Abelian geometric phases\cite{anandanNonadiabaticNonabelianGeometric1988,zanardiHolonomicQuantumComputation1999}. Their uniqueness lies in that the acquisition of phases depends only on the global characteristics of the quantum system's evolutionary trajectory, regardless of the specifics of the evolution process, thereby can enhance resistance to local noise\cite{zhuGeometricQuantumGates2005}.

Early schemes based on geometric phases relied on the system's adiabatic evolution\cite{wuGeometricPhaseGates2013}, which made quantum gates susceptible to environmental decoherence due to long gate times under adiabatic conditions. Therefore, the concept of non-adiabatic geometric quantum computing emerged\cite{wangNonadiabaticGeometricQuantum2007,sjoqvistNonadiabaticHolonomicQuantum2012}, aimed at shortening the operation time of quantum gates. Non-adiabatic geometric quantum computation is divided into two categories according to the geometric phase used: Nonadiabatic holonomic quantum computation(NHQC) for non-Abelian phases and Nonadiabatic geometric quantum computation(NGQC) for Abelian phases. Both types of quantum gates have undergone significant theoretical development\cite{liuPlugandplayApproachNonadiabatic2019,xuNonadiabaticHolonomicQuantum2012,liuNonadiabaticNoncyclicGeometric2020,sunOnestepImplementationRydberg2022,tangFastEvolutionSingle2022,liangCompositeShortpathNonadiabatic2022,liDynamicallyCorrectedNonadiabatic2021,dridiOptimalRobustQuantum2020} and experimental verification\cite{aiExperimentalRealizationNonadiabatic2020,xuExperimentalImplementationUniversal2020,fengExperimentalRealizationNonadiabatic2013,abdumalikovExperimentalRealizationNonAbelian2013} in recent years.

The global nature of quantum geometric phases makes geometric quantum computation a powerful strategy for advancing quantum computing towards large-scale and fault-tolerant development. However, despite NGQC's significant theoretical advantages, its robustness to control errors has not met expectations in practice. When subjected to control errors, its performance is similarly impacted as that of traditional dynamic schemes\cite{thomasRobustnessSinglequbitGeometric2011,zhengComparisonSensitivitySystematic2016}. In response to this challenge, although several schemes have been proposed to enhance the robustness of NGQC, they often increase gate operation time as a trade-off\cite{WOS:000824587200007,dingDynamicalcorrectedNonadiabaticGeometric2023,liuSuperrobustNonadiabaticGeometric2021}, which to some extent limits the potential application of the NGQC method in practical quantum computing.

To solve this problem, based on the traditional ``Super Robust" (SR) NGQC method\cite{liuSuperrobustNonadiabaticGeometric2021}, we propose a new NGQC optimization strategy called Fast Super Robust NGQC(FSR-NGQC), which not only maintains the robustness of NGQC gates to control errors but also significantly increases the speed of quantum gate operations, especially in phase rotation gates. Our method breaks the trade-off between robustness and operation speed in traditional NGQC schemes. By designing quantum evolutionary paths and pulse sequences, we have developed quantum gates that are resistant to control errors and ensure high-speed operations, offering a viable solution for efficient and robust quantum computing.

This paper is structured as follows. Chapter 2 introduces the traditional NGQC methods and existing schemes to enhance the robustness of NGQC. In Chapter 3, starting from the existing schemes, we extend to our designed FSR-NGQC scheme and provide the parameters for some main quantum gates. Chapter 4 first examines the robustness of our scheme against control errors. The analysis further extends to the perfomance of quantum gates under the dual effects of decoherence environments and control errors. By comparing with existing optimization strategies\cite{dingDynamicalcorrectedNonadiabaticGeometric2023}, we demonstrate the advantage of FSR-NGQC in resisting decoherence, which stems from its shorter gate times than traditional schemes. Lastly, we numerically calculate the fidelity performance of the two-qubit Control-T gate in the Rydberg atom system\cite{saffmanQuantumInformationRydberg2010,saffmanQuantumComputingAtomic2016,liuNonadiabaticNoncyclicGeometric2020}, the result of calculations prove the advantage of our scheme.
\section{Conventional Geometric Quantum Gate}
In this chapter, we will firstly introduce the conventional method of constructing a nonadiabatic geometric quantum gate\cite{chenNonadiabaticGeometricQuantum2018}, and then we introduce super robust NGQC scheme\cite{liuSuperrobustNonadiabaticGeometric2021}, which is a strategy that can suppress control errors to the fourth order. 
\subsection{Nonadiabatic Geometric Quantum Gate}
Consider an \(N\)-dimensional Hilbert space spanned by \(\{\left| \psi_k(0)\right\rangle\}_{k=1}^N\), where \(\left| \psi_k(t)\right\rangle\) are a set of orthonormal bases evolving according to the Schrödinger equation \(i\left| \dot{\psi}_k(t)\right\rangle =H(t)\left| \psi_k(t)\right\rangle\) (here we set \(\hbar=1\)). Then, the evolution equation of \(\left| \psi_k(t) \right\rangle\) can be written as \(\left| \psi_k(t) \right\rangle =U(t)\left| \psi_k(0) \right\rangle\), where \(U(t) = T\exp\left(-i\int_{0}^{t} H(t')dt'\right) = \sum_k \left| \psi_k(t) \right\rangle\left\langle \psi_k(0) \right|\) is the evolution operator, with \(T\) being the time-ordering operator, means that the integral is ordered in time. Construct a set of orthonormal auxiliary states \(\left| \mu_k(t) \right\rangle\) such that the space spanned by auxiliary states is equal to original space \(\{\left| \mu_k(t) \right\rangle\}_{k=1}^N = \{\left| \psi_k(t) \right\rangle\}_{k=1}^N\), then \(\left| \psi_k(t) \right\rangle\) can be expanded as \(\left| \psi_k(t) \right\rangle = \sum_{l=1}^N C_{lk}(t)\left| \mu_l(t) \right\rangle\), where \(C_{lk}(t)\) are time-dependent coefficients. Substituting this expansion into the Schrödinger equation yields

\[\dot{C}_{lk}(t) = i\sum_{m=1}^N [A_{lm}(t) - K_{lm}(t)] C_{mk}(t)\tag{1},\]
leading to the solution 
\[C(t) = T e^{i \int_0^t A\left(t^{\prime}\right)-K\left(t^{\prime}\right) \mathrm{d} t^{\prime}}\tag{2},\]
where \(A_{lm} \equiv i\left\langle \mu_l(t) \middle| \frac{\partial}{\partial t} \middle| \mu_m(t) \right\rangle\), \(K_{lm}(t) \equiv \left\langle \mu_l(t) \middle| H(t) \middle| \mu_m(t) \right\rangle\). If we require \(\left| \mu_k(0) \right\rangle = \left| \psi_k(0) \right\rangle\), then \(U(t)\) can be written as \(U(t) = \sum_k \left| \psi_k(t) \right\rangle\left\langle \mu_k(0) \right|\). Replacing \(\left| \psi_k(t) \right\rangle\) with the expansion in terms of auxiliary states, we get \(U(t) = \sum_{l,k=1}^N C_{lk}(t)\left| \mu_l(t) \right\rangle\left\langle \mu_k(0) \right|\). Incorporating the periodic evolution condition, which requires \(\left| \mu_k(\tau) \right\rangle = \left| \mu_k(0) \right\rangle\), the evolution operator can finally be written as 

\[U(\tau)=\sum_{l . k=1}^N\left[\mathcal{T} e^{i \int_0^\tau A\left(t^{\prime}\right)-K\left(t^{\prime}\right) \mathrm{d} t^{\prime}}\right]_{l k} \Pi_{l k}(0)\tag{3},\]
where \(\Pi_{lk}(0) \equiv \left| \mu_l(0) \right\rangle\left\langle \mu_k(0) \right|\).
To construct a general NGQC (Non-adiabatic Geometric Quantum Computation) gate using this evolution operator, for simplification purposes, we wish to avoid the situation where the matrices in the evolution operator are non-commutative, as this would complicate the analysis (requiring the use of time-ordering operators). we can require that \(A_{lk}(t) - K_{lk}(t) = 0\) for \(l \neq k\), this simplifies the evolution operator to 

\[U(\tau) = \sum_{l=1}^N e^{i\int_0^\tau \left(A_{ll}(t') - K_{ll}(t')\right)dt'} \Pi_{ll}(0)\tag{4}.\]

Further, to ensure that the evolution operator is related only to the geometric phase, that is, only related to \(A_{lm}\), the accumulation of dynamic phase must be zero. by requiring \(\int_0^\tau K_{ll}(t')dt' = 0\), the evolution operator is determined entirely by the geometric phase, denoted as 

\[U(\tau,0) = e^{i\gamma_1} \left| \mu_1(0) \right\rangle \left\langle \mu_1(0) \right| + e^{i\gamma_2} \left| \mu_2(0) \right\rangle \left\langle \mu_2(0) \right| \label{5}\tag{5},\]
where \(\gamma_1\) and \(\gamma_2\) represent the geometric phases accumulated over the evolution for each states. This formulation ensures that the overall phase of the quantum gate is purely geometric, eliminating any contribution from dynamic phases. When we choose specific auxiliary states \(\left| \mu_1(t) \right\rangle\) and \(\left| \mu_2(t) \right\rangle\) as follows:

\[
\begin{aligned}
	\left| \mu_1(t) \right\rangle &= \cos\left(\frac{\theta(t)}{2}\right)\left| 0 \right\rangle + \sin\left(\frac{\theta(t)}{2}\right)e^{i\phi(t)}\left| 1 \right\rangle, \\
	\left| \mu_2(t) \right\rangle &= \sin\left(\frac{\theta(t)}{2}\right)e^{-i\phi(t)}\left| 0 \right\rangle - \cos\left(\frac{\theta(t)}{2}\right)\left| 1 \right\rangle,
\end{aligned} \label{6}\tag{6}
\]
both \(\theta(t)\) and \(\phi(t)\) are time-dependent parameters with a period \(\tau\), leading to the result \(\gamma_2 = -\gamma_1\). The evolution operator can be further simplified to \(U(\tau) = U_n (\gamma) = e^{i\gamma \mathbf{n} \cdot \boldsymbol{\sigma}}\), where \(\mathbf{n} = (\sin \theta_0 \cos \phi_0, \sin \theta_0 \sin \phi_0, \cos \theta_0)\) signifies a rotation around the axis \(\mathbf{n}\), then we can construct arbitrary single-qubit quantum gate. 
\subsection{Super Robust NGQC}
However, such geometric quantum gates do not possess a significant advantage in terms of resistance to control errors. Under the same conditions of control error interference, the fidelity loss experienced by geometric quantum gates is on the same order of magnitude as that of dynamic gates\cite{thomasRobustnessSinglequbitGeometric2011}. To fully exploit the advantages of geometric quantum gates, various strategies have been developed to reduce control errors\cite{WOS:000824587200007,dingDynamicalcorrectedNonadiabaticGeometric2023}. Predecessors has mathematically summarized the conditions under which geometric quantum gates (including NGQC and NHQC) achieve a reduction of control errors to the fourth order (whereas traditional schemes are of the second order). NGQC that meet these criteria are termed SR-NGQC\cite{liuSuperrobustNonadiabaticGeometric2021}. Below, the conditions for SR-NGQC are introduced.

Considering a control error \(\epsilon\), which is typically small relative to the Hamiltonian and can be viewed as a perturbation. The Hamiltonian with control error is denoted as \(H'(t) = (1 + \epsilon)H(t)\), and the evolution operator becomes \(U_E(\tau) = T\exp\left(-i\int_0^{\tau} H'(t)dt\right)\). For this time integral, taking a Magnus series expansion\cite{WOS:000263313300001}, the formula is transformed into \(U_E(\tau) = \exp\left(\sum_{k=1}^{\infty} \Lambda_k(\tau)\right)\), where \(\Lambda_k(\tau)\) are the terms of the Magnus series. Expanding it at \(\epsilon\) and calculating its fidelity, the relation between fidelity and \(\epsilon\) can be expressed as:

\[
\begin{aligned}
	F &= \frac{1}{M} \left| \text{Tr}\left(U_E U_C^\dagger \right) \right| \\
	&\approx 1 - \frac{\epsilon^2}{2M} \sum_{m=1, k=1}^M |D_{km}|^2 - O(\epsilon^4),\\
\end{aligned}\tag{7}
\]
where \(D_{km} \equiv \int_0^{\tau} \langle \psi_k(t) | H(t) | \psi_m(t) \rangle dt\). Considering the evolution operator introduced above in EQ.(\ref{5}), when \(k=m\), \(D_{mm} \equiv \int_0^{\tau} \langle \mu_m(t) | H(t) | \mu_m(t) \rangle dt\) is the same requirement as for the dynamical phase being zero. For \(k \neq m\), since \(|\psi_m(t)\rangle = e^{i\gamma_m} |\mu_m(t)\rangle\), then \(D_{km} = \int_0^{\tau} e^{i(\gamma_m - \gamma_k)} A_{km} dt\). Meeting this condition can endow NGQC with fourth-order robustness against control errors.

\section{Fast Super Robust NGQC}
In this chapter, we derive our scheme from SR-NGQC, resulting in what we call FSR-NGQC. This version not only maintains resistance to control errors at the fourth order but also enhances the operational speed of quantum gates. Below, we present our derivation process.

The traditional SR-NGQC requires a longer time duration of \(3\pi / \Omega_0\)\cite{WOS:000996294300003}. It is longer than the time required for traditional NGQC, which just require \(2\pi / \Omega_0\). The longer gate duration increases the system's sensitivity to decoherence, making the design of fast and high-fidelity NGQC of highly importance. Following the traditional SR-NGQC concept, we have designed FSR-NGQC scheme that maintains fourth-order robustness to control errors while reducing the gate time to less than \(3\pi / \Omega_0\) (see TABLE.\ref{tab:1}), which we will introduce below.

\begin{figure}
	\centering
	\captionsetup{justification=raggedright,singlelinecheck=false}
	\begin{subfigure}[b]{0.45\linewidth}
		\includegraphics[width=1\textwidth,keepaspectratio]{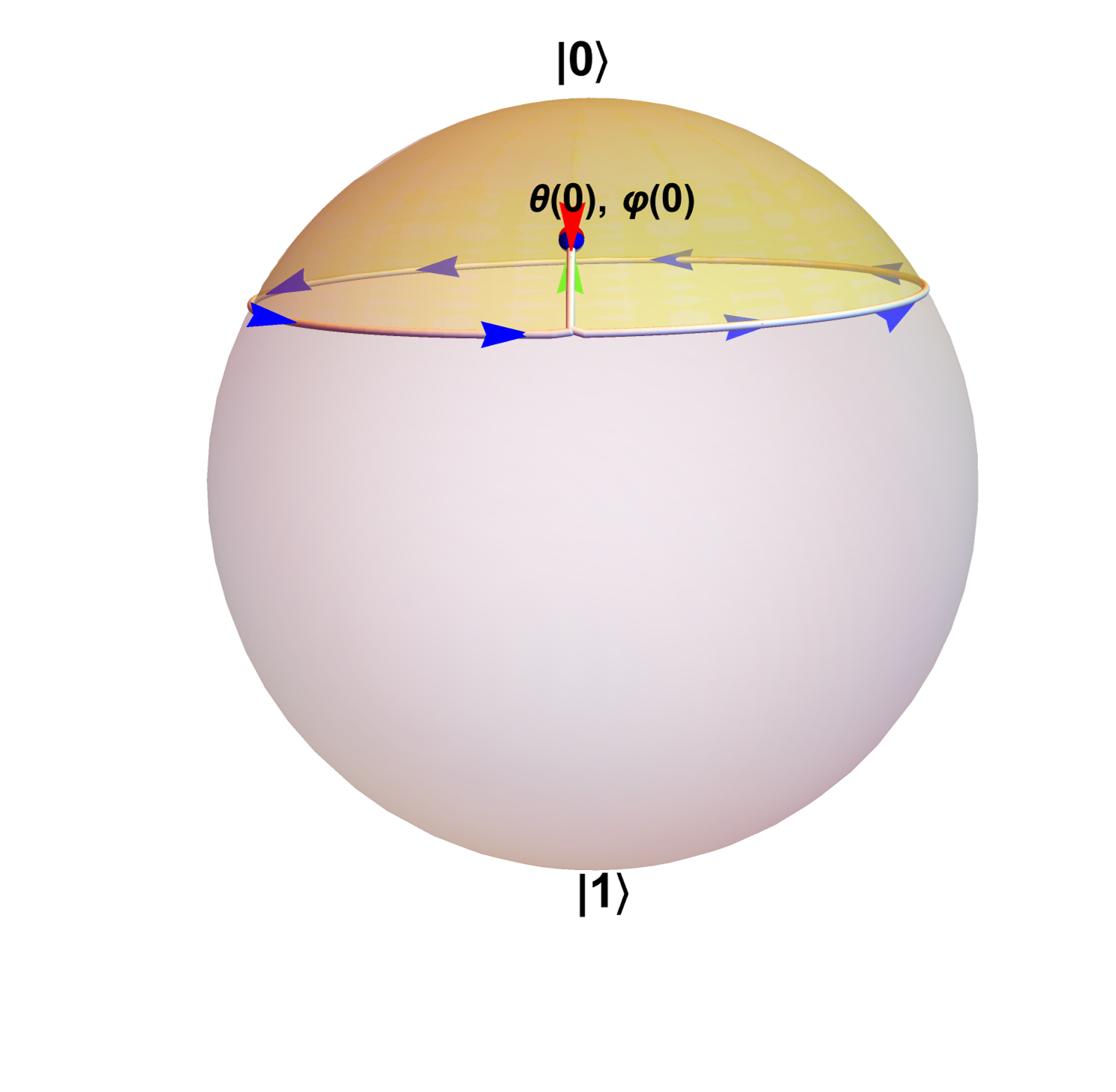}%
		\caption{\label{1}}
		\label{fig:sub-a}
	\end{subfigure}
	\begin{subfigure}[b]{0.45\linewidth}
		\includegraphics[width=1\textwidth,keepaspectratio]{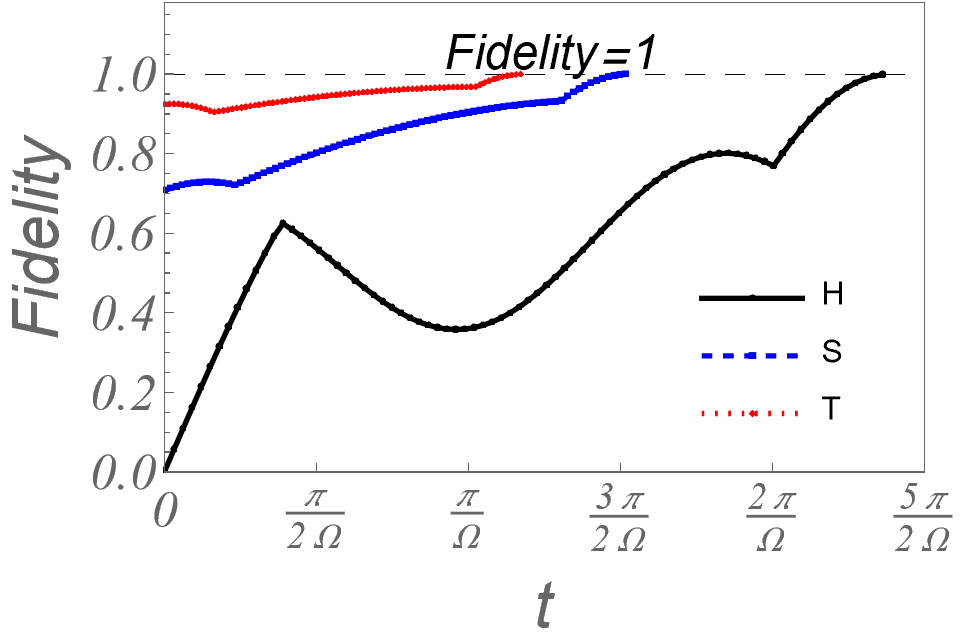}%
		\caption{\label{2}}
		\label{fig:sub-b}
	\end{subfigure}
	\caption{Evolution of FSR-NGQC Gate. (a) The evolution path of the states $|\mu_1(t)\rangle$ on the Bloch sphere starts from the red point and follows the red arrow. It moves to a certain position before it begins rotating around the z-axis. After completing the rotation, it returns to the starting point along the original path, which is indicated by the green arrow. (b) The evolution of the fidelity for the $H$, $S$, and $T$ gates using our scheme is shown. Notably, our scheme requires less time than $3\pi$. } 
	\label{fig:1}
\end{figure}

Considering the computational space \(\{|0\rangle, |1\rangle\}\) and using a set of standard orthogonal auxiliary states as EQ.(\ref{6}), we find that \(-A_{22}(t) = A_{11}(t) = \sin^2(\theta(t)/2) \dot{\phi}(t)\), and \(A_{12}(t) = A_{21}^*(t) = -\frac{1}{2} e^{i\phi(t)} [\dot{\theta}(t) + i\sin(\theta(t))\dot{\phi}(t)]\) where \(\dot{\phi}(t)\) represents \(\partial\phi(t)/\partial t\) and \(\dot{\theta}(t)\) represents \(\partial\theta(t)/\partial t\). By employing the inverse engineering method\cite{daemsRobustQuantumControl2013}, we construct Hamiltonian as:
\[
\begin{aligned}
	H(t) = & \left[ A_{12}(t) |\mu_1(t)\rangle \langle \mu_2(t)| + \text{H.c.} \right]\\
	& +K_{11}(t)|\mu_1(t)\rangle \langle \mu_1(t)| + K_{22}(t)|\mu_2(t) \rangle \langle \mu_2(t)|.\\
\end{aligned}\tag{8}
\]
Thereby, the system evolution satisfies the condition \(A_{lk}(t) - K_{lk}(t) = 0\) for \(l \neq k\). When meeting the condition for periodic evolution, the system's evolution operator is same as EQ.(\ref{5}). Where \(\gamma_m = \int_0^{\tau} A_{mm}(t) - K_{mm}(t) dt\), the super-robust condition can be written as:
\[
\begin{aligned}[t]
	D_{12} &= \int_0^{\tau} \langle \psi_1(t) | H(t) | \psi_2(t) \rangle dt\\
	&= \int_0^{\tau} \langle \mu_1(t) | e^{-i\gamma_1(t)} H(t) e^{i\gamma_2(t)} | \mu_2(t) \rangle dt\\
	&= \int_0^{\tau} -\frac{1}{2} e^{-i[\gamma_1(t) - \gamma_2(t)]} e^{i\phi(t)} [\dot{\theta}(t) + i \sin(\theta(t)) \dot{\phi}(t)] dt\\
	&=\int_0^{\tau} -\frac{1}{2} e^{-i\int_0^t 2 A_{11}(t')+K_{22}(t')-K_{11}(t') dt'}\\        
	&\;\;\;\;\cdot e^{i\phi(t)}[\dot{\theta}(t)+i\sin(\theta(t))\dot{\phi}(t)]dt.\\
\end{aligned}\tag{9}
\]
It can be seen that this form is an integral nested within another integral, making its analytic solution highly complex, But if we set \(K_{11}(t) = -K_{22}(t) = A_{11}(t) - \dot{\beta}(t)\), and further impose \(\beta(0) = 0\), then the super-robust condition becomes:
\[
D_{12} = \int_0^{\tau} -\frac{1}{2} e^{-2i\beta(t)} e^{i\phi(t)} [\dot{\theta}(t) + i \sin(\theta(t)) \dot{\phi}(t)] dt = 0.\tag{10}
\]
By choosing a simple \(\beta(t)\), this integral condition becomes solvable. It is important to note that \(\beta(t)\) must satisfy the condition for the cancellation of dynamical phase, \(\int_0^{\tau} K_{ll}(t') dt' = 0\), leading to \(\beta(\tau) = \int_0^{\tau} A_{11}(t') dt'\), which is similar to \(\gamma\) considering only the geometric phase. It equals the geometric phase at the end of the evolution but experiences different changes along the path.

In this paper, we divide the Hamiltonian into three segments, with the functions \(\beta(t)\), \(\theta(t)\), and \(\phi(t)\) each having different forms between these segments. Specifically, we set the Hamiltonian as:
\[
H(t) = \frac{1}{2} \begin{pmatrix}
	\Delta(t) & \Omega(t)e^{-i\psi(t)} \\
	\Omega(t)e^{i\psi} & -\Delta(t)
\end{pmatrix}.\tag{11}
\]
according to inverse engineering schemes\cite{daemsRobustQuantumControl2013}, we acquire these equations: \(\Delta(t) = 2[\cos(\theta(t)) \dot{\beta}(t) + \sin^2(\theta(t)/2) \dot{\phi}(t)]\), \(\Omega(t) =\sqrt{[\sin(\theta(t))\cdot[2\dot{\beta}(t) - \dot{\phi}(t)]]^2 + (\dot{\theta}(t))^2} \), \(\psi(t)=\exp\left\{i\left[\arctan\left(\frac{\dot{\theta}(t)}{\sin(\theta(t))\cdot[2\dot{\beta}(t)-\dot{\phi}(t)]}\right)-\phi(t)\right]\right\}\). We set \(\beta(t)\), \(\theta(t)\), and \(\phi(t)\) as lines with different slopes in the three segments, satisfying the condition that boundary values are equal, as specified in TABLE.\ref{tab:1}. 
\begin{table*}[t]
	\centering
	\begin{tabular}{|l|l|l|l|}
		\hline
		\diagbox{Config}{Time} & $0 \leq t \leq t_1$ & $t_1<t \leq t_2$ & $t_2<t<\tau$ \\
		\hline
		$\theta(t)$ & $a_1 t+b_1$ & $a_1 t_1+b_1$ & $a_2\left(t-t_2\right)+a_1 t_1+b_1$ \\
		\hline
		$\varphi(t)$ & $c_1$ & $c_1+c_2\left(t-t_1\right)$ & $c_1+c_2\left(t_2-t_1\right)$ \\
		\hline
		$\beta(t)$ & $d_1 t$ & $d_1 t_1+d_2\left(t-t_1\right)$ & $d_1 t_1+d_2\left(t_2-t_1\right)+d_3\left(t-t_2\right)$ \\
		\hline
	\end{tabular}
	\caption{Shape of functions \(\beta(t)\), \(\theta(t)\), and \(\phi(t)\). For simplicity, all these functions are designed to be linear. }
	\label{tab:1}
\end{table*}
The evolution path of auxiliary state $|\mu_1(t)\rangle$ on the Bloch sphere is shown in FIG.\ref{fig:1}. Starting from the red point and following the red arrow, it moves to a certain position before starting to rotate around the z-axis. After rotating back, it returns to the starting point along the original path marked by the green arrow. The half of solid angle swept by the angle of rotation around the z-axis exactly equals the acquired geometric phase \(\gamma\). By selecting \(U[\gamma_1, \theta(0), \phi(0)]\), any single-qubit quantum gate can be constructed. Here, we demonstrate the parameter settings for the $H$, $S$, and $T$ gates (see TABLE.\ref{tab:2}). The $H$ gate requires \(U[\pi/2, \pi/2, 0]\), for $S$ and $T$ gates are \(U[\pi/4, 0, 0]\) and \(U[\pi/8, 0, 0]\), respectively. By denoting the absolute value of \(D_{12}\) as a loss function, the parameter values satisfying the super-robust condition can be numerically obtained using a gradient descent optimization algorithm\cite{ruder2017overview}. The closer the value of \(D_{12}\) is to 0, the better the effect of eliminating second-order control errors. We present the selected parameter values for these three gates(see TABLE.\ref{tab:2}), and the fidelity of these three gates over time (FIG.\ref{fig:1}), where the pulse intensity \(|\Omega| \leq \Omega_{\text{Max}}\). It can be seen that the operation time for all three gates is less than that required by traditional SR-NGQC, which is \(3\pi/\Omega\).

\section{Gate application and perfomance}
In this chapter, we first prove that our scheme is robust against control errors. Then, through numerical calculations, we demonstrate that our scheme exhibits superior fidelity performance under decoherence effects compared to existing schemes. Finally, we illustrate and calculate the fidelity performance of the two-qubit Control-T gate in a Rydberg atom system. By comparing our scheme with the DC\cite{dingDynamicalcorrectedNonadiabaticGeometric2023} and SI\cite{WOS:000936540400002} schemes, our approach consistently shows advantages.
\subsection{Gate perfomance under control error}
\begin{figure}[p]
	\captionsetup{justification=raggedright,singlelinecheck=false}
	\begin{subfigure}[b]{0.45\linewidth}
		\includegraphics[width=\linewidth]{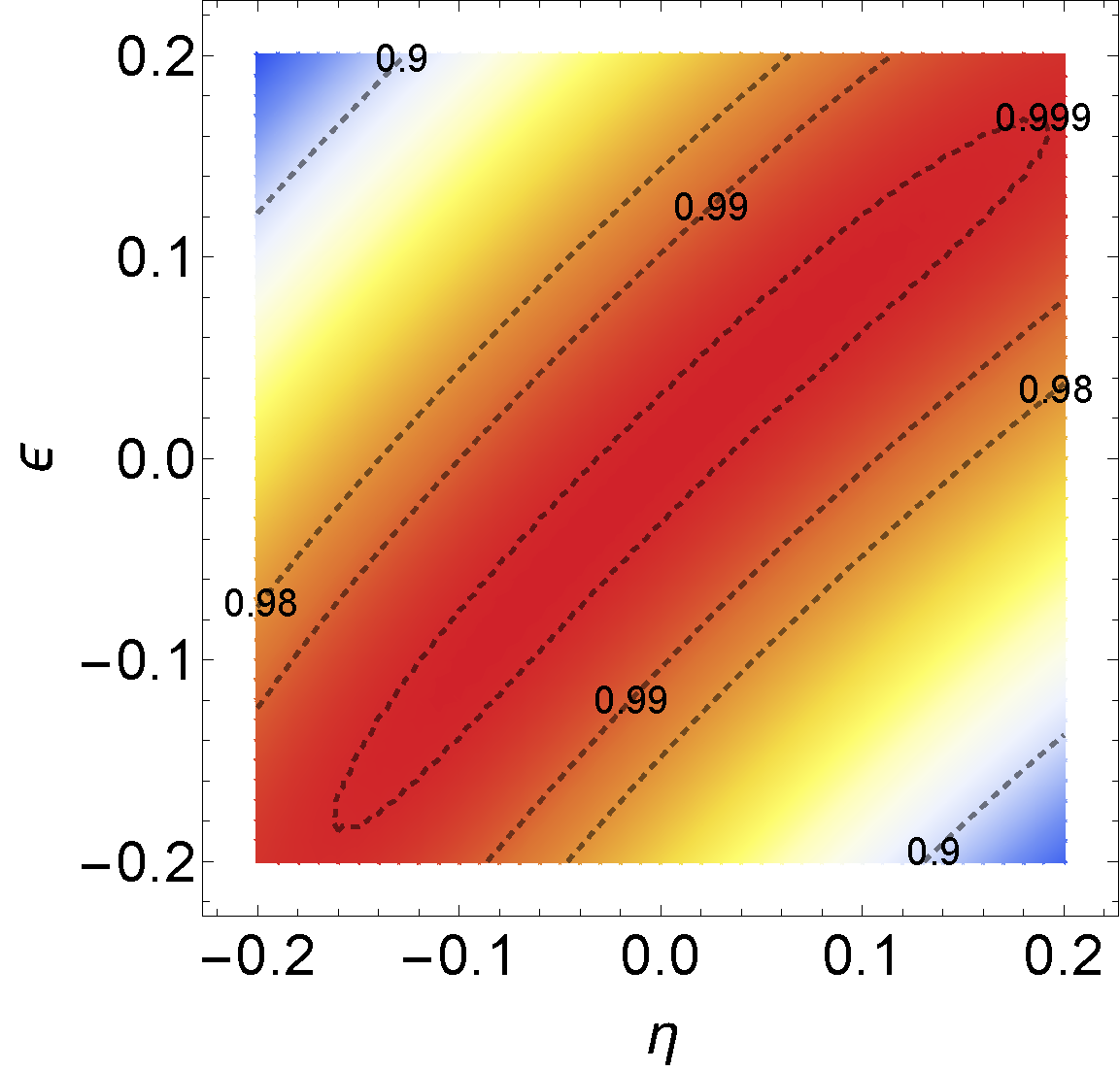}
		\caption{} 
		\label{fig:sub-a}
	\end{subfigure}
	\begin{subfigure}[b]{0.45\linewidth}
		\includegraphics[width=\linewidth]{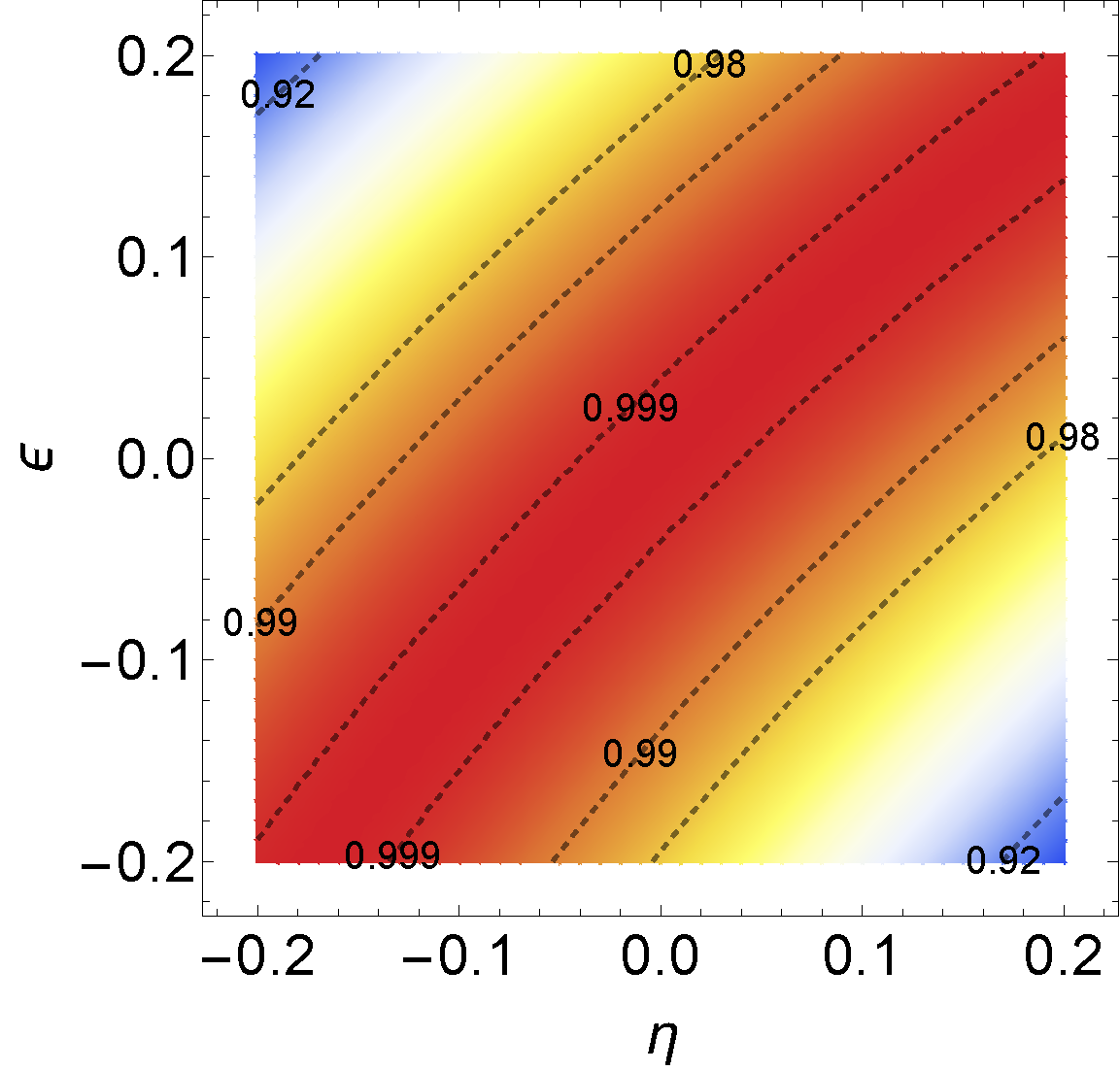}
		\caption{} 
		\label{fig:sub-b}
	\end{subfigure}
	\begin{subfigure}[b]{0.45\linewidth}
		\includegraphics[width=\linewidth]{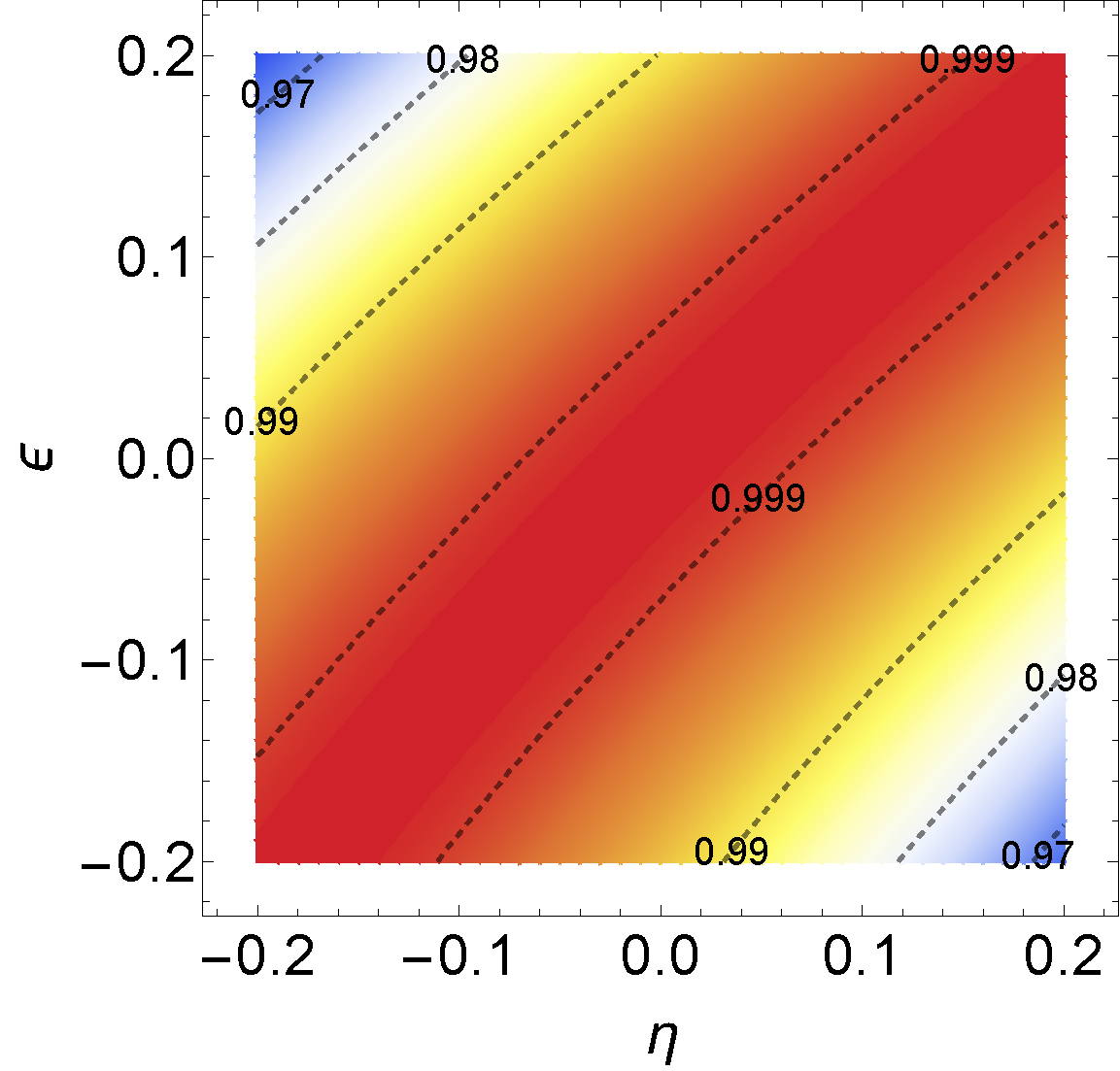}
		\caption{} 
		\label{fig:sub-c}
	\end{subfigure}
	\begin{subfigure}[b]{0.45\linewidth}
		\includegraphics[width=\linewidth]{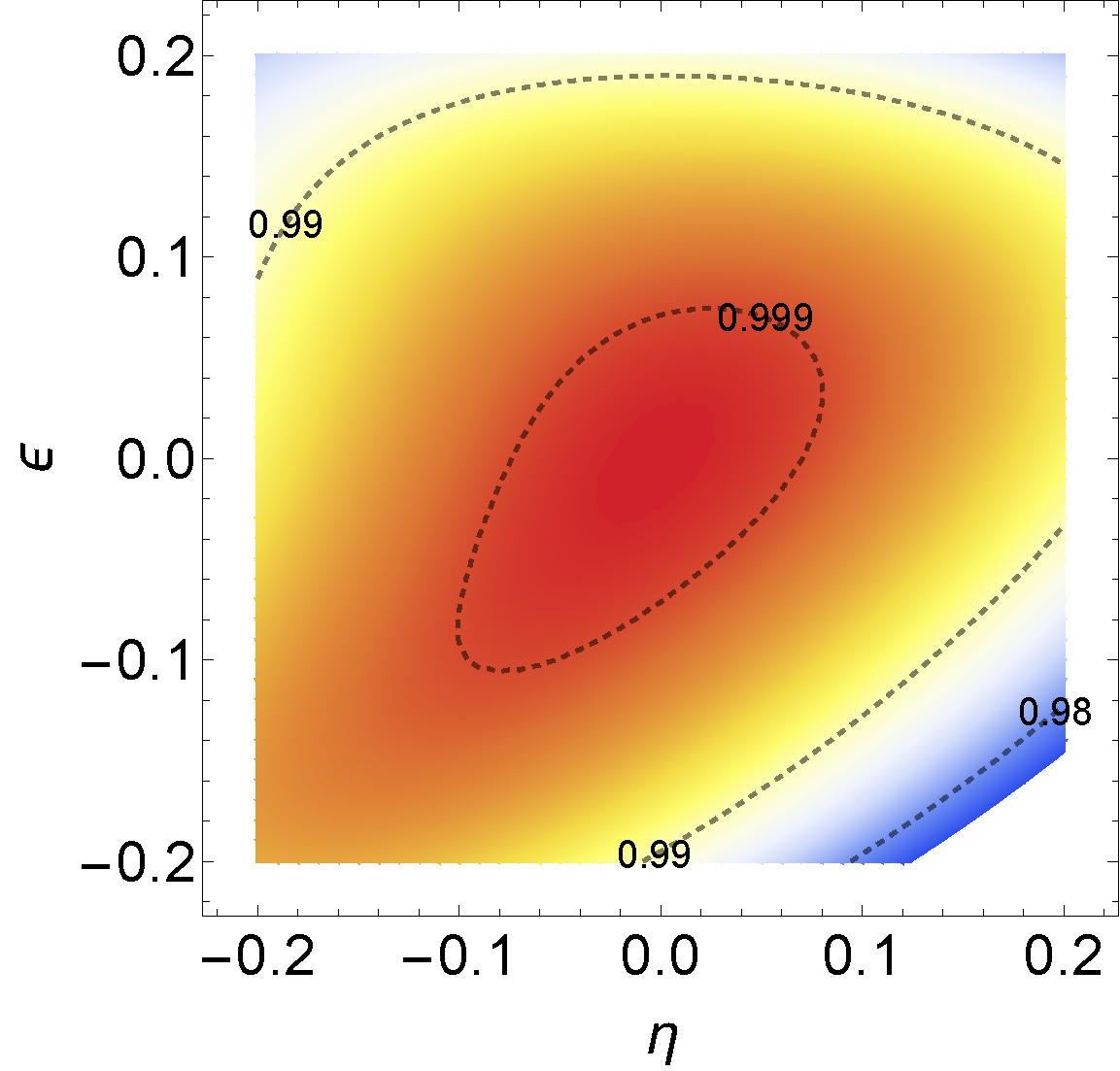}
		\caption{} 
		\label{fig:sub-d}
	\end{subfigure}
	\begin{subfigure}[b]{1\linewidth}
		\includegraphics[width=\linewidth]{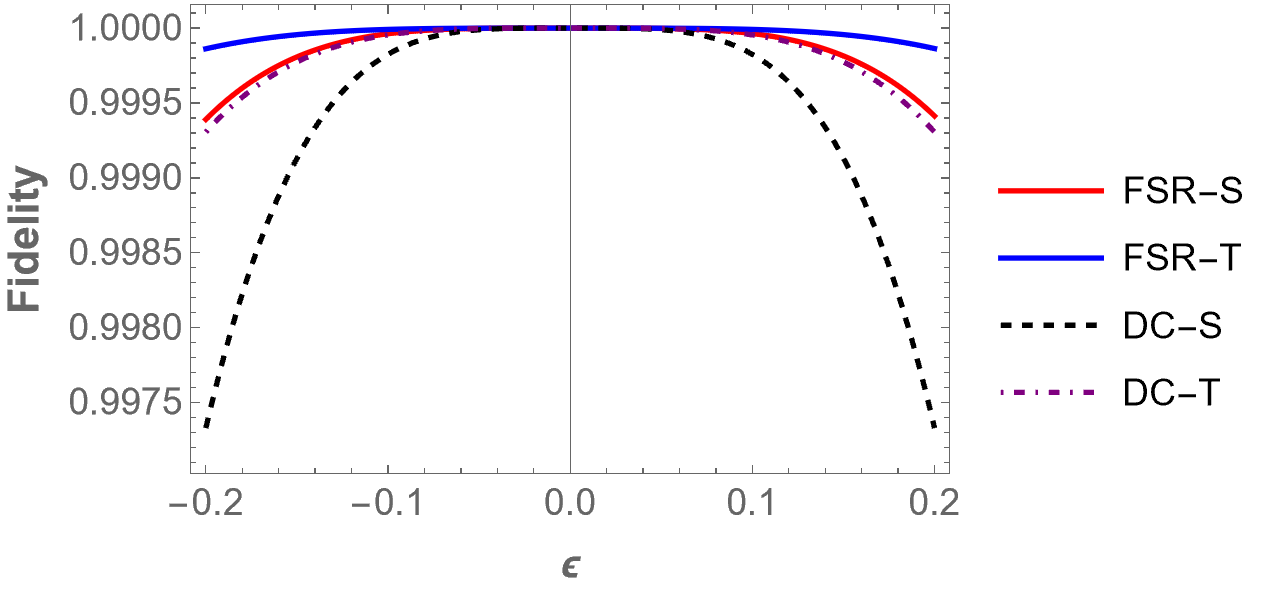}
		\caption{} 
		\label{fig:sub-e}
	\end{subfigure}
	\caption{Figure demonstrates the fidelity under control error for (a) $H$, (b) $S$, and (c) $T$ gates using the FSR-NGQC scheme, (d) the H gate using the DC-NGQC scheme. It is evident that our scheme achieves a fidelity comparable to that of the DC scheme, assuming $\eta$= $\epsilon$} 
	\label{fig:2}
\end{figure}
We first discuss the fidelity performance of gates under control errors. Consider the Hamiltonian with error as follows:
\[ H^E(t) = \frac{1}{2} \begin{pmatrix} (1+\eta)\Delta(t) & (1+\epsilon)\Omega(t)e^{-i\phi(t)} \\ (1+\epsilon)\Omega(t)e^{i\phi(t)} & -(1+\eta)\Delta(t) \end{pmatrix},\tag{12} \]
where \(\eta, \epsilon\) range from \([-0.2, 0.2]\). Without considering other error factors (such as decoherence), we plotted the fidelity graphs of $H$, $S$, and $T$ gates under different error parameter values. Considering that no literature has yet provided parameters for implementing these three gates using SR-NGQC, we compare our scheme with DC-NGQC. It can be observed that under the condition of \(\eta = \epsilon\), the FSR-NGQC gates exhibit good robustness to control errors (comparable to DC-NGQC, quantum gates with fourth-order control error robustness, as shown in FIG.\ref{fig:2}), which is exactly what the super-robust condition guarantees. However, when \(\eta\) and \(\epsilon\) are inversely proportional, the global error condition considered by the super robust condition fails, hence the poor robustness performance of the quantum gates, which is a shortcoming that needs to be addressed in the future. Considering the case of \(\eta = \epsilon\), we plotted the robustness performance graph of $S$ and $T$ gates compared to DC-NGQC under control errors, showing that our scheme has higher robustness than DC-NGQC. When considering decoherence effects, our scheme will also perform better because the pulse area of our $H$, $S$, $T$ gates is smaller than the approximately \(4\pi / \Omega\) required by DC-NGQC, which will be demonstrated in later chapters.
\begin{table*}[t]
	\centering
	\caption{Gate Parameters. We set \(\Omega_{\text{Max}} = 1\). The pulse area is calculated by the integral \(\int_0^\tau \Omega(t) \, dt\). It is important to note that in our scheme, the pulse $\Omega$ does not always reach its maximum point.}
	\label{tab:2}
	\begin{tabular}{|l|l|l|l|l|l|l|l|l|l|l|l|l|}
		\hline 
		Parameter & $a_1$ & $b_1$ & $a_2$ & $c_1$ & $c_2$ & $d_1$ & $d_2$ & $d_3$ & $t_1$ & $t_2$ & $\tau$ & Pulse Area \\
		\hline 
		H Gate & 0.2159 & 0.7854 & -0.2159 & \;\;0\; & 1.2348 & 0.5637 & 0.0400 & 0.5637 & 1.2123 & 6.3005 & $2.4 \pi$ & $2.32\pi$ \\
		\hline 
		S Gate & 0.9857 & 0.0000 & -0.9857 & \;\;0\; & 1.8678 & 0.1273 & 0.1779 & 0.1273 & 0.7332 & 4.0971 & $1.53\pi$ & $1.53\pi$ \\
		\hline 
		T Gate & 0.9987 & 0.0000 & -0.9987 & \;\;0\; & 2.3161 & 0.0521 & 0.1252 & 0.0521 & 0.5060 & 3.2187 & $1.18\pi$ & $1.18\pi$ \\
		\hline
	\end{tabular}
\end{table*}
\subsection{Gate perfomance under decoherence}
\begin{figure}[pt]
	
	\captionsetup{justification=raggedright,singlelinecheck=false}
	\begin{subfigure}[b]{0.45\linewidth}
		\includegraphics[width=\linewidth]{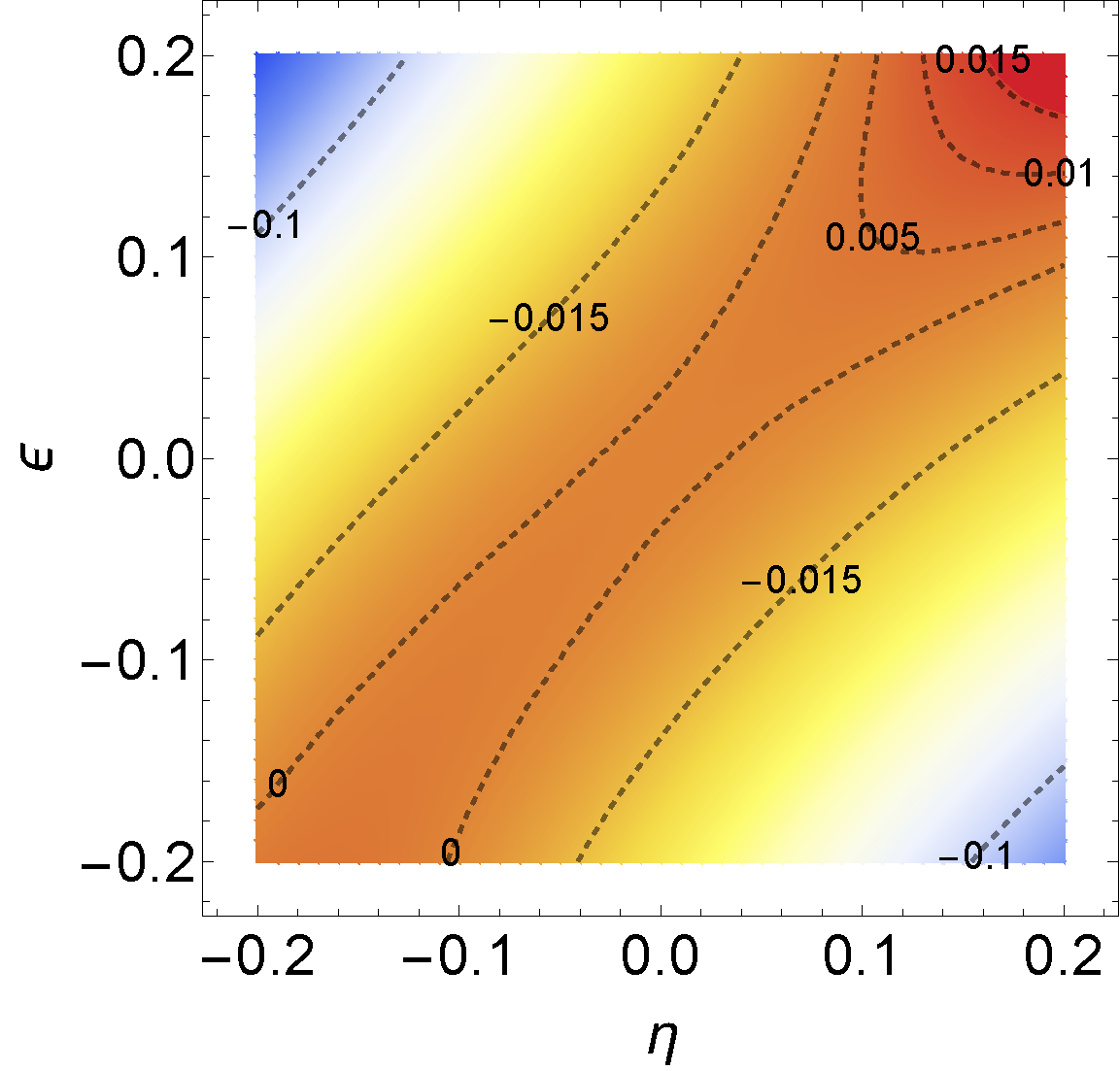}
		\caption{} 
		\label{fig:sub-a}
	\end{subfigure}
	\begin{subfigure}[b]{0.45\linewidth}
		\includegraphics[width=\linewidth]{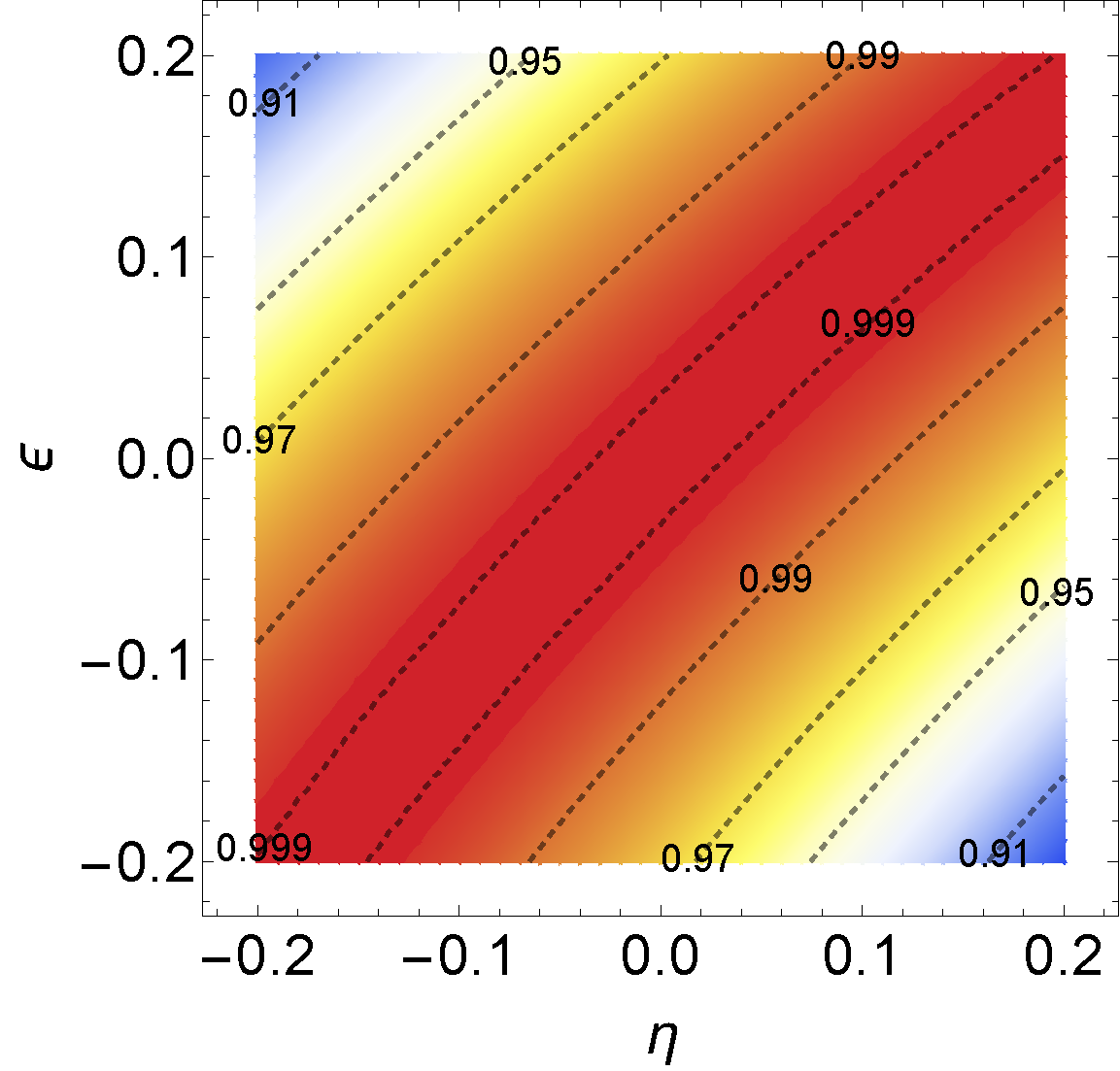}
		\caption{} 
		\label{fig:sub-b}
	\end{subfigure}
	
	\begin{subfigure}[b]{0.45\linewidth}
		\includegraphics[width=\linewidth]{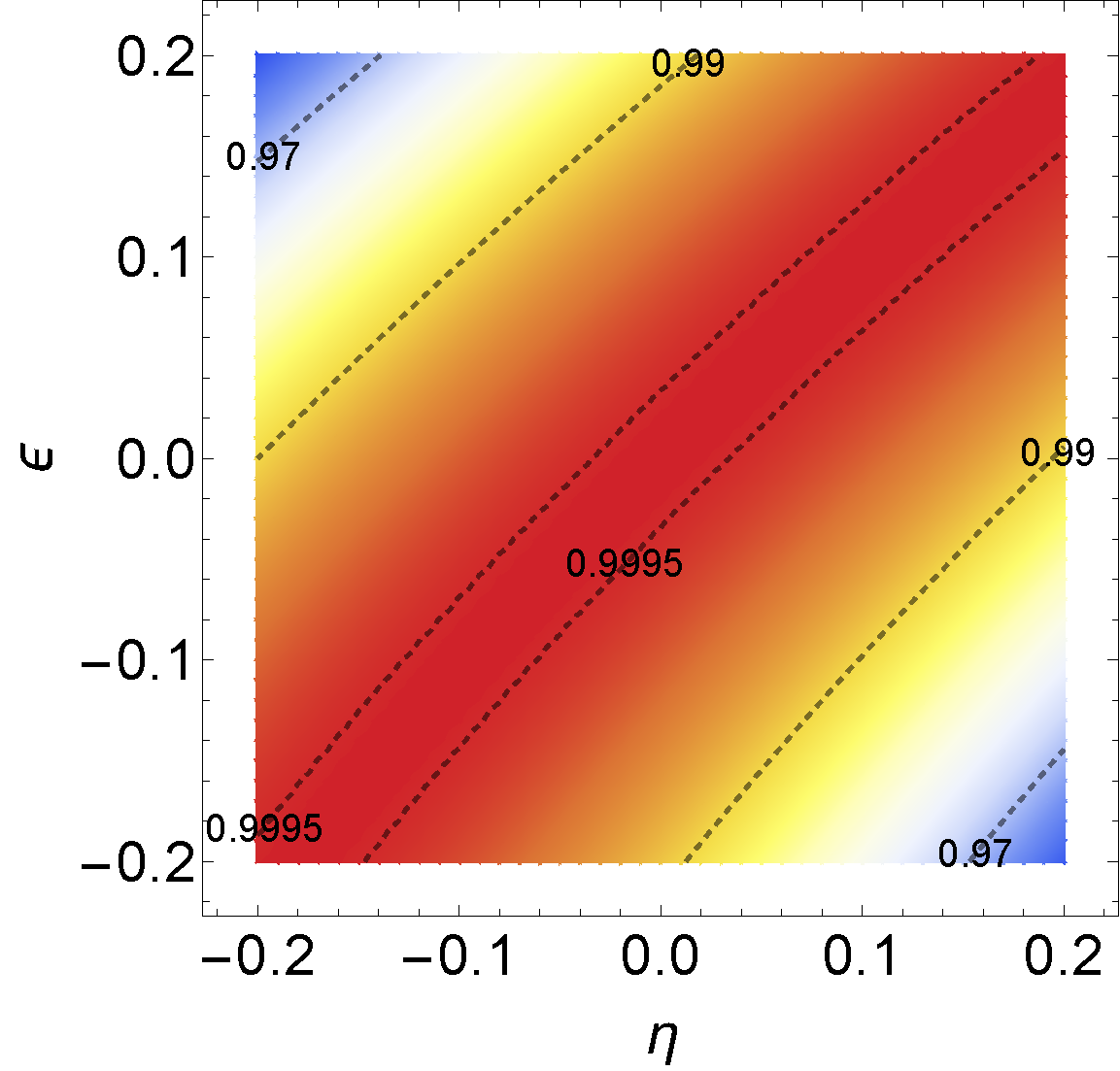}
		\caption{} 
		\label{fig:sub-c}
	\end{subfigure}
	\begin{subfigure}[b]{0.45\linewidth}
		\includegraphics[width=\linewidth]{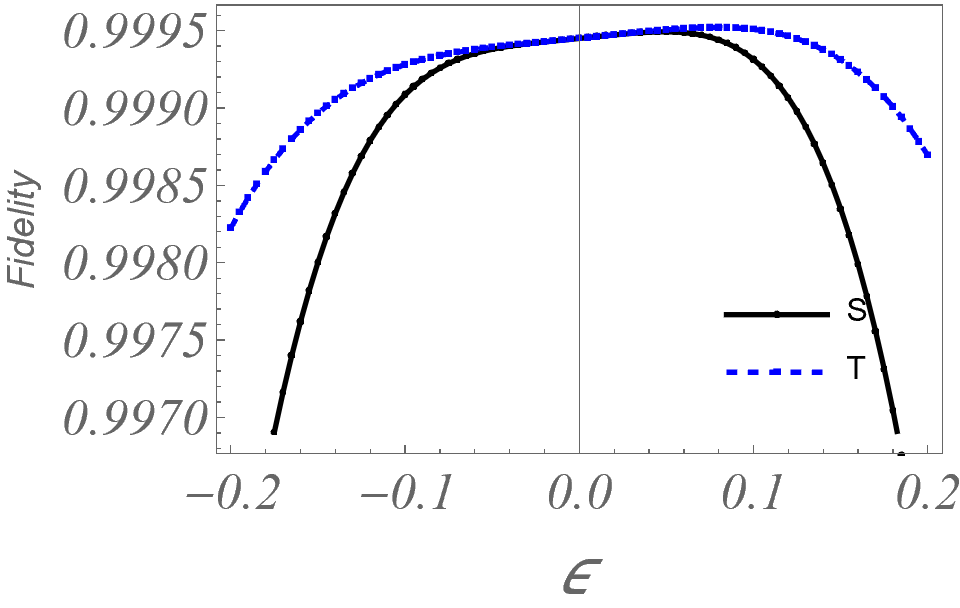}
		\caption{} 
		\label{fig:sub-d}
	\end{subfigure}
	\caption{Performance under decoherence. Figure (a) depicts the fidelity difference for the H gate between our scheme and the DC-NGQC scheme, while figure (b) and (c) show the performance of the S and T gates under decoherence in our scheme, respectively. Figure (d) illustrates the performance of the S and T gates under decoherence in the DC scheme. It can be observed that, when the globality of control errors is effective, our scheme outperforms the DC scheme under decoherence and control error.} 
	\label{fig:3}
\end{figure}
In practical situations, quantum gates often suffer from decoherence, which is one of the main challenges faced by quantum computers. To simulate the performance of our scheme in a real physical model, we choose the Rydberg atom system\cite{saffmanQuantumInformationRydberg2010} for demonstration. Following the setup in the article\cite{liangStateIndependentNonadiabaticGeometric2023}, where logical qubits are encoded in two hyperfine ground states \(|0\rangle \equiv |5S_{1/2}, F=1, m_F=0\rangle\) and \(|1\rangle \equiv |5S_{1/2}, F=2, m_F=0\rangle\). controlled through two-photon Raman transitions\cite{saffmanQuantumComputingAtomic2016}, we can achieve the Hamiltonian model required by our scheme. The performance of the quantum gates is calculated using the master equation method\cite{lindbladGeneratorsQuantumDynamical1976}, with the formula:
\[
\dot{\rho}(t) = i[\rho(t), H(t)] + \frac{1}{2} [\Gamma_1 L(\sigma_1) + \Gamma_2 L(\sigma_2)],\tag{13}
\]
where \(L(\sigma) = 2\sigma\rho\sigma^\dagger - \sigma^\dagger \sigma\rho - \rho\sigma^\dagger \sigma\), \(\Gamma_1\) and \(\Gamma_2\) are the decay coefficient and dephasing coefficient, respectively. In our chosen physical model, the coupling strength is \(\Omega = 2\pi \times 1.36MHz\), and the decoherence rate is \(\Gamma = 1.15kHz \approx \Omega / 7400\), with only a dephasing coefficient present. At this time, \(\sigma_2 = |1\rangle\langle1| - |0\rangle\langle0|\). We select four initial states \(\{|0\rangle, |1\rangle, (|0\rangle + |1\rangle)/\sqrt{2}, (|0\rangle + i|1\rangle)/\sqrt{2}\}\) to calculate the average fidelity, and define fidelity as \(F = (1/4) \sum_{k=1}^4 \langle \Psi_k(\tau) | \rho_k(\tau) | \Psi_k(\tau) \rangle\), where \(\rho(\tau)\) is the final evolved density matrix, and \(\Psi(\tau)\) is the expected state. Similarly, we consider \(\eta\) and \(\epsilon\) values in the range \([-0.2, 0.2]\) and plot the graph of fidelity versus the two types of errors (FIG.\ref{fig:3}). For $S$($T$) gate, When \(\eta\) equals \(\epsilon\), the fidelity remains above 99.9\%(99.95\%), whereas under the same conditions, the traditional DC-NGQC drop to 99.7\%(99.85\%). This is because the pulse area of the S and T gates in our FSR-NGQC scheme is smaller than that of the traditional scheme, and the faster gate speed reduces the effects of decoherence.
\subsection{2-Qubit Gate}

Next, we examine a 2-qubit quantum gate, still utilizing the setup outlined in\cite{liangStateIndependentNonadiabaticGeometric2023}. Consider two rubidium atoms, one serving as the control qubit and the other as the target qubit. In addition to the two fine-structure ground states used in single-qubit gates, we introduce an auxiliary state \(|r\rangle \equiv |83S, J=\frac{1}{2}, m_J=\frac{1}{2}\rangle\). The \(|0\rangle\) state of the control qubit is resonantly coupled to the \(|r\rangle\) state with Rabi frequency \(\Omega_1(t)\), and the \(|1\rangle\) state of the target qubit is non-resonantly coupled to the \(|r\rangle\) state with Rabi frequency \(\Omega_2(t)\) and detuning \(\Delta\), along with a precisely controllable Rydberg-Rydberg interaction term. The Hamiltonian can be written as:
\[
\begin{aligned}
	H(t) = &[\Omega_1(t)|r\rangle_1 \langle0| + \text{H.c.}] + [\Omega_2(t)e^{-i\Delta t} |r\rangle_2 \langle1| + \text{H.c.}]\\
	& + V_{12} |rr\rangle_{12} \langle rr| + \eta\Omega' [|r\rangle_1 \langle r| + |r\rangle_2 \langle r|],
\end{aligned}\tag{14}
\]
where \(\eta\) represents the detuning error, and \(\Omega'\) is the magnitude of \(\Omega_2\). When the condition \(V_{12} \gg \{\bar{\Omega}_1, \Omega'\}\) is satisfied, where \(\bar{\Omega}_1\) is the absolute value of \(\Omega_1\), and \(\bar{\Omega}_1 \gg \Omega'\), the effective Hamiltonian can be written as:
\[
H_{\text{eff}}(t) = |1\rangle_1 \langle1| \otimes H_2(t).\tag{15}
\]
When transformed into the interaction picture with \(e^{-i\Delta/2(|1r\rangle_{12} \langle1r| - |11\rangle_{12} \langle11|)}\), the effective Hamiltonian becomes:
\[
\begin{aligned}
	H_{\text{eff}}^I(t) =&\ [\Omega_2(t)|11\rangle_{12} \langle1r| + \text{H.c.}]\\
	&\ + \frac{\Delta}{2} [|11\rangle_{12} \langle11| - |1r\rangle_{12} \langle1r|],
\end{aligned}\tag{16}
\]
with this form of the Hamiltonian, it is possible to apply an arbitrary phase to the \(|11\rangle\) state. Here, we demonstrate a Controlled-T (C-T) gate, with a phase magnitude of \(\pi/4\), to fully showcase the advantage of our scheme in precisely controlling phases.

\[ \text{C-T} = \begin{pmatrix} 1 & 0 & 0 & 0 \\ 0 & 1 & 0 & 0 \\ 0 & 0 & 1 & 0 \\ 0 & 0 & 0 & e^{i\pi/4} \end{pmatrix}. \tag{17}\]

\begin{figure}[t]
	\captionsetup{justification=raggedright,singlelinecheck=false}
	\begin{subfigure}[b]{0.45\linewidth}
		\includegraphics[width=1\textwidth,keepaspectratio]{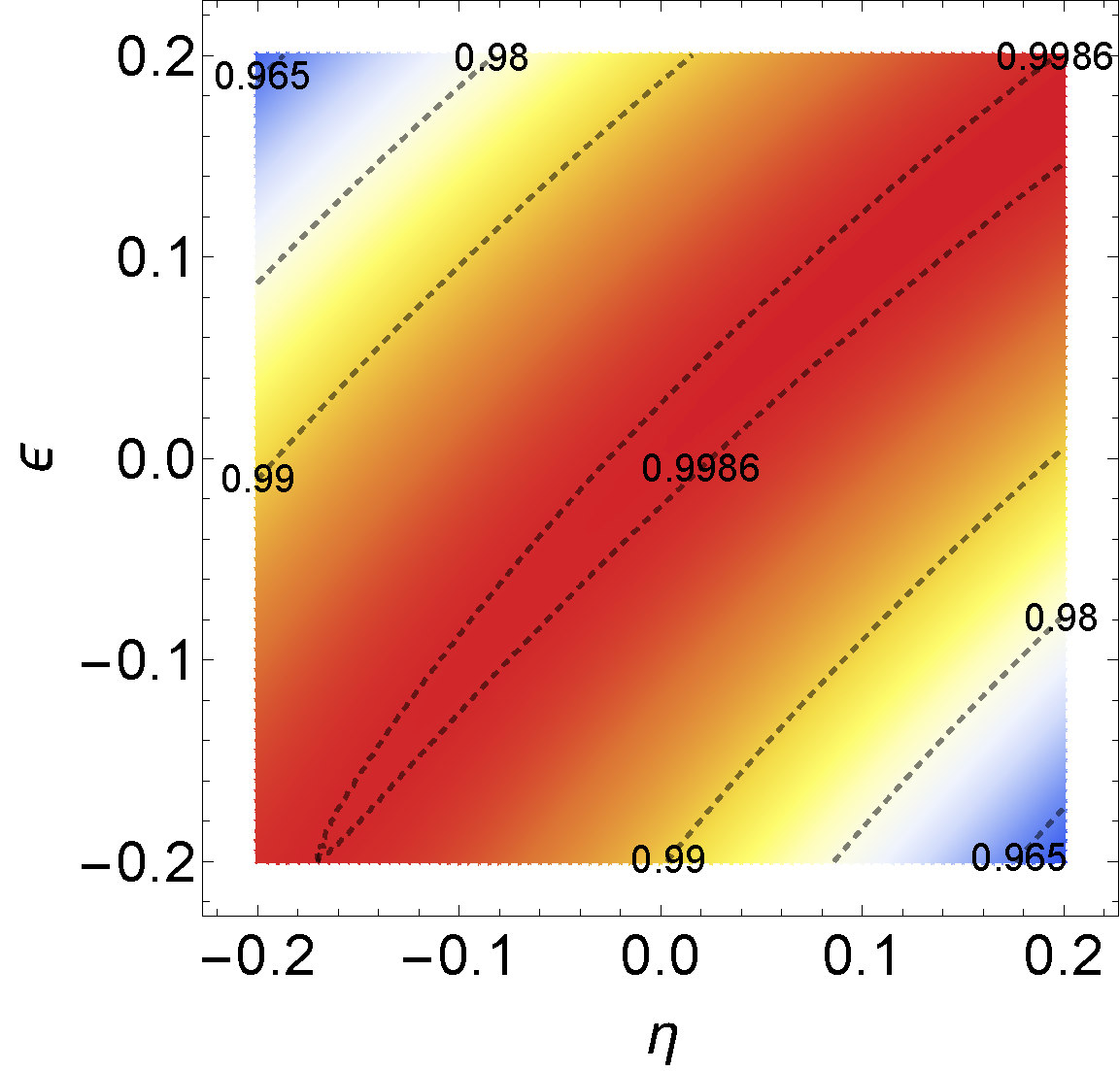}%
		\caption{\label{1}}
		\label{fig:sub-a}
	\end{subfigure}
	\begin{subfigure}[b]{0.45\linewidth}
		\includegraphics[width=1\textwidth,keepaspectratio]{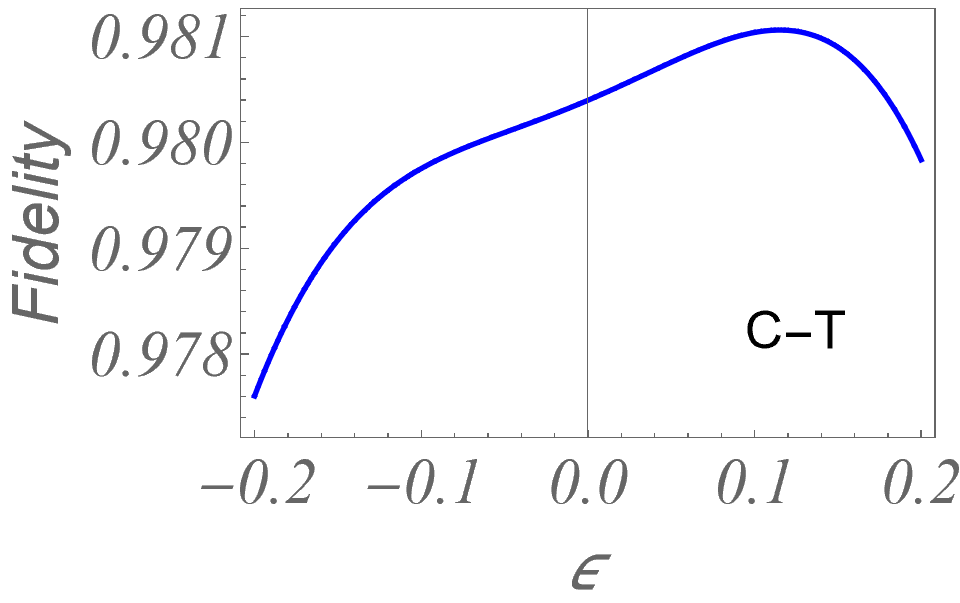}%
		\caption{\label{2}}
		\label{fig:sub-b}
	\end{subfigure}
	\begin{subfigure}[b]{0.8\linewidth}
		\includegraphics[width=1\textwidth,keepaspectratio]{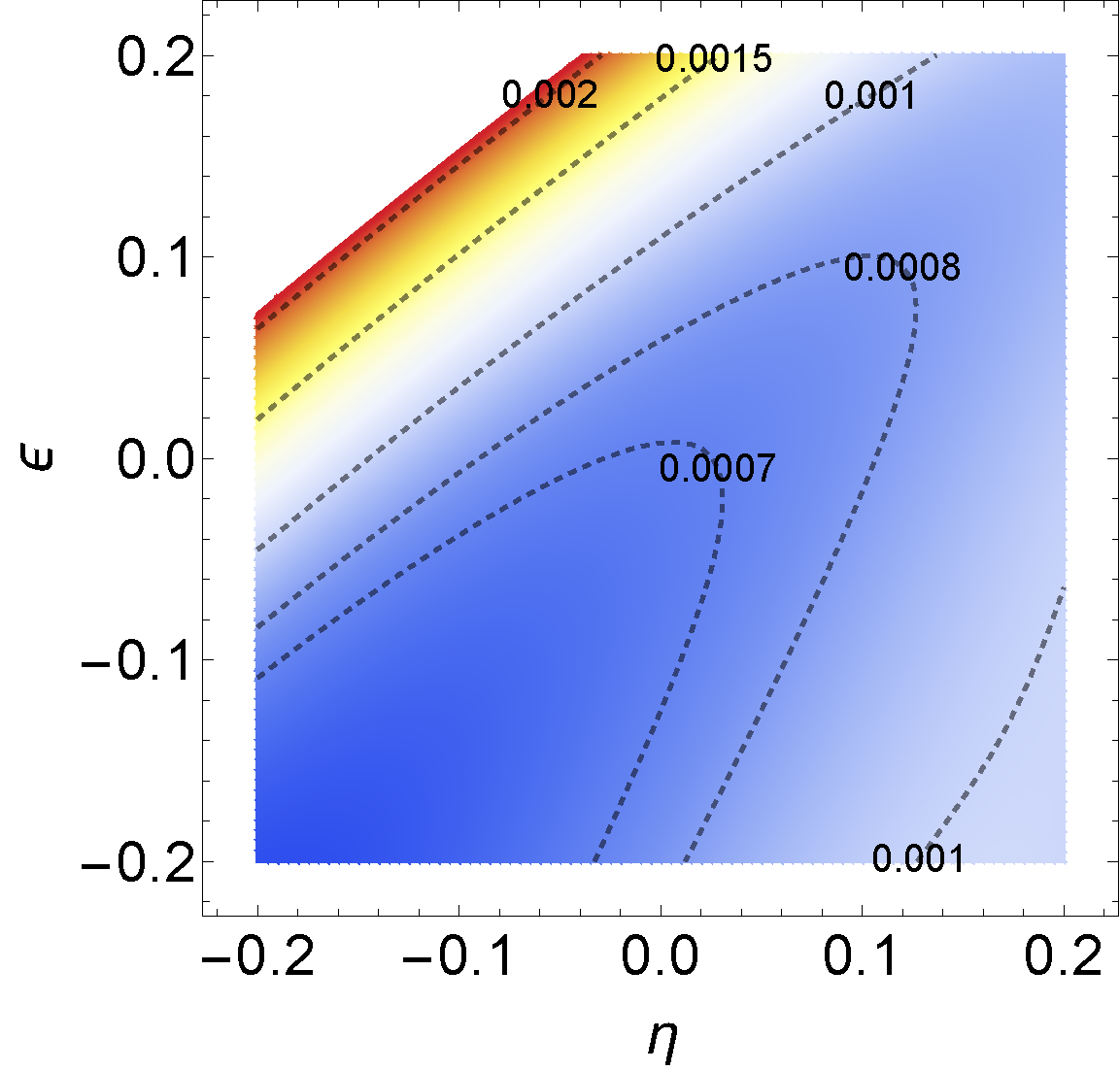}%
		\caption{\label{3}}
		\label{fig:sub-c}
	\end{subfigure}
	\caption{2bit Gate under decoherence. Figure(a) shows the fidelity of C-T gate using FSR scheme under decoherence and control error, figure(b) shows the fidelity of C-T gate using DC scheme. Our scheme can achieve a fidelity of up to 99.86\%, whileas DC scheme can only achieve 98.1\%, (c) illustrates the fidelity difference between our scheme and the SI scheme under decoherence. It can be observed that our scheme generally achieves higher fidelity. } 
	\label{fig:4}
\end{figure}

We adhere to the previous definition of fidelity, and since the auxiliary level \(|r\rangle\) has been introduced, we must consider the decay rate from this level to all ground states, with the rate \(\Gamma = 1/\tau_r\), the inverse of the Rydberg state lifetime \(\tau_r\)(uniformly taken as 50\(\mu\)s). Other ground states are denoted as garbage states \(|2\rangle\), and we can write the two-qubit gate's decoherence operators as: \(\sigma_2^0 = |0\rangle_2 \langle r|\), \(\sigma_2^1 = |1\rangle_2 \langle r|\), \(\sigma_2^2 = |2\rangle_2 \langle r|\), where we set \(\Gamma_2^0 = \Gamma_2^1 = \Gamma/8\), \(\Gamma_2^2 = 3\Gamma/4\), assuming decay probabilities from Rydberg state to all ground states is equal. Adding the previously considered dephasing operator \(\sigma = |1\rangle\langle1| - |0\rangle\langle0|\), we calculated the fidelity of the two-qubit gate under this decoherence environment as a function of control errors, noting that the coupling strength chosen is \(\Omega' = 2\pi \times 0.75MHz\). From FIG.\ref{fig:4}, it can be seen that our two-qubit gate scheme still performs well under decoherence. With \(\eta\) and \(\epsilon\) being close or equal, it achieves a very high fidelity (99.88\%). In contrast, traditional schemes suffer from poorer performance in a decoherence environment due to longer gate operation times. The DC scheme only achieves a maximum fidelity of 98.1\%, and in the worst case, the fidelity drops to 97.76\%. Our scheme also exhibits superiority over similar optimization strategies like SI-NGQC\cite{liangStateIndependentNonadiabaticGeometric2023}. We have plotted the fidelity difference under decoherence, and it is evident that our scheme achieves higher fidelity across most regions. This advantage is attributed to our scheme's equivalent level of control error robustness, combined with greater speed (our C-T gate requires only \(1.53\pi\), compared to \(1.61\pi\) required by the SI scheme).

\section{Conclusion and outlook}
In conclusion, based on the previous SR-NGQC theory, we have developed a new fast and robust FSR-NGQC scheme by relaxing the constraint for the elimination of dynamic phases. We provide complete parameters for FSR $H$, $S$, and $T$ gates and achieves faster speeds with similar robustness for implementing phase rotation gates compared to previous NGQC schemes
, thus performing well in a decoherence environment. We also presented a 2-qubit Control-T gate implementation scheme in a Rydberg atom system, characterized by high robustness and speed, maintaining high fidelity even with short lifetimes of Rydberg states. Therefore, we offer an effective scheme for realizing fast and robust geometric quantum gates.

For further extensions, although the current speed increase is not substantial, by considering different forms of the $\beta$ parameter, we could possibly obtain even better performance. Our scheme could also be combined with decoherence-free subspaces\cite{xuNonadiabaticHolonomicQuantum2012,sunOnestepImplementationRydberg2022} (DFS) to construct quantum gates robust against global dephasing.

A limitation of our approach is the introduction of detuning, where the robustness can decrease when the control errors in detuning and coupling strength differ, as the premise of SR theory—globality of control errors—gets violated. We demonstrated this through fidelity variation with control errors graphs, pointing to the need for a more refined theory to eliminate control errors.



\bibliographystyle{elsarticle-num} 
\bibliography{321}

\begin{thebibliography}{10}
\expandafter\ifx\csname url\endcsname\relax
  \def\url#1{\texttt{#1}}\fi
\expandafter\ifx\csname urlprefix\endcsname\relax\def\urlprefix{URL }\fi
\expandafter\ifx\csname href\endcsname\relax
  \def\href#1#2{#2} \def\path#1{#1}\fi

\bibitem{preskillQuantumComputingNISQ2018}
J.~Preskill, Quantum {{Computing}} in the {{NISQ}} era and beyond, QUANTUM 2
  (Aug. 2018).
\newblock \href {https://doi.org/10.22331/q-2018-08-06-79}
  {\path{doi:10.22331/q-2018-08-06-79}}.

\bibitem{bhartiNoisyIntermediatescaleQuantum2022}
K.~Bharti, A.~{Cervera-Lierta}, T.~H. Kyaw, T.~Haug, S.~{Alperin-Lea},
  A.~Anand, M.~Degroote, H.~Heimonen, J.~S. Kottmann, T.~Menke, W.-K. Mok,
  S.~Sim, L.-C. Kwek, A.~{Aspuru-Guzik}, Noisy intermediate-scale quantum
  algorithms, REVIEWS OF MODERN PHYSICS 94~(1) (Feb. 2022).
\newblock \href {https://doi.org/10.1103/RevModPhys.94.015004}
  {\path{doi:10.1103/RevModPhys.94.015004}}.

\bibitem{higgottVariationalQuantumComputation2019}
O.~Higgott, D.~Wang, S.~Brierley, Variational {{Quantum Computation}} of
  {{Excited States}}, QUANTUM 3 (Jul. 2019).
\newblock \href {https://doi.org/10.22331/q-2019-07-01-156}
  {\path{doi:10.22331/q-2019-07-01-156}}.

\bibitem{tillyVariationalQuantumEigensolver2022}
J.~Tilly, H.~Chen, S.~Cao, D.~Picozzi, K.~Setia, Y.~Li, E.~Grant, L.~Wossnig,
  I.~Rungger, G.~H. Booth, J.~Tennyson, The {{Variational Quantum
  Eigensolver}}: {{A}} review of methods and best practices, PHYSICS
  REPORTS-REVIEW SECTION OF PHYSICS LETTERS 986 (2022) 1--128.
\newblock \href {https://doi.org/10.1016/j.physrep.2022.08.003}
  {\path{doi:10.1016/j.physrep.2022.08.003}}.

\bibitem{zhouQuantumApproximateOptimization2020}
L.~Zhou, S.-T. Wang, S.~Choi, H.~Pichler, M.~D. Lukin, Quantum {{Approximate
  Optimization Algorithm}}: {{Performance}}, {{Mechanism}}, and
  {{Implementation}} on {{Near-Term Devices}}, PHYSICAL REVIEW X 10~(2) (Jun.
  2020).
\newblock \href {https://doi.org/10.1103/PhysRevX.10.021067}
  {\path{doi:10.1103/PhysRevX.10.021067}}.

\bibitem{eganFaulttolerantControlErrorcorrected2021}
L.~Egan, D.~M. Debroy, C.~Noel, A.~Risinger, D.~Zhu, D.~Biswas, M.~Newman,
  M.~Li, K.~R. Brown, M.~Cetina, C.~Monroe, Fault-tolerant control of an
  error-corrected qubit, NATURE 598~(7880) (2021) 281--+.
\newblock \href {https://doi.org/10.1038/s41586-021-03928-y}
  {\path{doi:10.1038/s41586-021-03928-y}}.

\bibitem{ciracQuantumComputationsCold1995}
{Cirac}, {Zoller}, Quantum {{Computations}} with {{Cold Trapped Ions}}.,
  Physical review letters 74~(20) (1995) 4091--4094.
\newblock \href {https://doi.org/10.1103/PhysRevLett.74.4091}
  {\path{doi:10.1103/PhysRevLett.74.4091}}.

\bibitem{clarkeSuperconductingQuantumBits2008}
J.~Clarke, F.~K. Wilhelm, Superconducting quantum bits, NATURE 453~(7198)
  (2008) 1031--1042.
\newblock \href {https://doi.org/10.1038/nature07128}
  {\path{doi:10.1038/nature07128}}.

\bibitem{golterOptomechanicalQuantumControl2016}
D.~A. Golter, T.~Oo, M.~Amezcua, K.~A. Stewart, H.~Wang, Optomechanical
  {{Quantum Control}} of a {{Nitrogen-Vacancy Center}} in {{Diamond}}, PHYSICAL
  REVIEW LETTERS 116~(14) (Apr. 2016).
\newblock \href {https://doi.org/10.1103/PhysRevLett.116.143602}
  {\path{doi:10.1103/PhysRevLett.116.143602}}.

\bibitem{levineHighFidelityControlEntanglement2018}
H.~Levine, A.~Keesling, A.~Omran, H.~Bernien, S.~Schwartz, A.~S. Zibrov,
  M.~Endres, M.~Greiner, V.~Vuletic, M.~D. Lukin, High-{{Fidelity Control}} and
  {{Entanglement}} of {{Rydberg-Atom Qubits}}, PHYSICAL REVIEW LETTERS 121~(12)
  (Sep. 2018).
\newblock \href {https://doi.org/10.1103/PhysRevLett.121.123603}
  {\path{doi:10.1103/PhysRevLett.121.123603}}.

\bibitem{yanTunableCouplingScheme2018}
F.~Yan, P.~Krantz, Y.~Sung, M.~Kjaergaard, D.~L. Campbell, T.~P. Orlando,
  S.~Gustavsson, W.~D. Oliver, Tunable {{Coupling Scheme}} for {{Implementing
  High-Fidelity Two-Qubit Gates}}, PHYSICAL REVIEW APPLIED 10~(5) (Nov. 2018).
\newblock \href {https://doi.org/10.1103/PhysRevApplied.10.054062}
  {\path{doi:10.1103/PhysRevApplied.10.054062}}.

\bibitem{niuUniversalQuantumControl2019}
M.~Y. Niu, S.~Boixo, V.~N. Smelyanskiy, H.~Neven, Universal quantum control
  through deep reinforcement learning, NPJ QUANTUM INFORMATION 5 (Apr. 2019).
\newblock \href {https://doi.org/10.1038/s41534-019-0141-3}
  {\path{doi:10.1038/s41534-019-0141-3}}.

\bibitem{zhangGeometricHolonomicQuantum2023}
J.~Zhang, T.~H. Kyaw, S.~Filipp, L.-C. Kwek, E.~Sjoqvist, D.~Tong, Geometric
  and holonomic quantum computation, PHYSICS REPORTS-REVIEW SECTION OF PHYSICS
  LETTERS 1027 (2023) 1--53.
\newblock \href {https://doi.org/10.1016/j.physrep.2023.07.004}
  {\path{doi:10.1016/j.physrep.2023.07.004}}.

\bibitem{berryQuantalPhaseFactors1984}
M.~Berry, Quantal phase factors accompanying adiabatic changes, Proceedings of
  the Royal Society of London, Series A (Mathematical and Physical Sciences)
  392~(1802) (1984) 45--57.
\newblock \href {https://doi.org/10.1098/rspa.1984.0023}
  {\path{doi:10.1098/rspa.1984.0023}}.

\bibitem{ekertGeometricQuantumComputation2000}
A.~Ekert, M.~Ericsson, P.~Hayden, H.~Inamori, {\relax JA}.~Jones, {\relax
  DKL}.~Oi, V.~Vedral, Geometric quantum computation, JOURNAL OF MODERN OPTICS
  47~(14-15) (2000) 2501--2513.
\newblock \href {https://doi.org/10.1080/09500340008232177}
  {\path{doi:10.1080/09500340008232177}}.

\bibitem{anandanNonadiabaticNonabelianGeometric1988}
J.~Anandan, Non-adiabatic non-abelian geometric phase, Physics Letters A
  133~(4-5) (1988) 171--5.
\newblock \href {https://doi.org/10.1016/0375-9601(88)91010-9}
  {\path{doi:10.1016/0375-9601(88)91010-9}}.

\bibitem{zanardiHolonomicQuantumComputation1999}
P.~Zanardi, M.~Rasetti, Holonomic quantum computation, PHYSICS LETTERS A
  264~(2-3) (1999) 94--99.
\newblock \href {https://doi.org/10.1016/S0375-9601(99)00803-8}
  {\path{doi:10.1016/S0375-9601(99)00803-8}}.

\bibitem{zhuGeometricQuantumGates2005}
{\relax SL}.~Zhu, P.~Zanardi, Geometric quantum gates that are robust against
  stochastic control errors, PHYSICAL REVIEW A 72~(2) (Aug. 2005).
\newblock \href {https://doi.org/10.1103/PhysRevA.72.020301}
  {\path{doi:10.1103/PhysRevA.72.020301}}.

\bibitem{wuGeometricPhaseGates2013}
H.~Wu, E.~M. Gauger, R.~E. George, M.~Mottonen, H.~Riemann, N.~V. Abrosimov,
  P.~Becker, H.-J. Pohl, K.~M. Itoh, M.~L.~W. Thewalt, J.~J.~L. Morton,
  Geometric phase gates with adiabatic control in electron spin resonance,
  PHYSICAL REVIEW A 87~(032326) (Mar. 2013).
\newblock \href {https://doi.org/10.1103/PhysRevA.87.032326}
  {\path{doi:10.1103/PhysRevA.87.032326}}.

\bibitem{wangNonadiabaticGeometricQuantum2007}
Z.~S. Wang, C.~Wu, X.-L. Feng, L.~C. Kwek, C.~H. Lai, C.~H. Oh, V.~Vedral,
  Nonadiabatic geometric quantum computation, PHYSICAL REVIEW A 76~(044303)
  (Oct. 2007).
\newblock \href {https://doi.org/10.1103/PhysRevA.76.044303}
  {\path{doi:10.1103/PhysRevA.76.044303}}.

\bibitem{sjoqvistNonadiabaticHolonomicQuantum2012}
E.~Sjoqvist, D.~M. Tong, L.~M. Andersson, B.~Hessmo, M.~Johansson, K.~Singh,
  Non-adiabatic holonomic quantum computation, NEW JOURNAL OF PHYSICS
  14~(103035) (Oct. 2012).
\newblock \href {https://doi.org/10.1088/1367-2630/14/10/103035}
  {\path{doi:10.1088/1367-2630/14/10/103035}}.

\bibitem{liuPlugandplayApproachNonadiabatic2019}
B.-J. Liu, X.-K. Song, Z.-Y. Xue, X.~Wang, M.-H. Yung, Plug-and-play approach
  to nonadiabatic geometric quantum gates, PHYSICAL REVIEW LETTERS 123~(100501)
  (Sep. 2019).
\newblock \href {https://doi.org/10.1103/PhysRevLett.123.100501}
  {\path{doi:10.1103/PhysRevLett.123.100501}}.

\bibitem{xuNonadiabaticHolonomicQuantum2012}
G.~F. Xu, J.~Zhang, D.~M. Tong, E.~Sjoqvist, L.~C. Kwek, Nonadiabatic holonomic
  quantum computation in decoherence-free subspaces, PHYSICAL REVIEW LETTERS
  109~(170501) (Oct. 2012).
\newblock \href {https://doi.org/10.1103/PhysRevLett.109.170501}
  {\path{doi:10.1103/PhysRevLett.109.170501}}.

\bibitem{liuNonadiabaticNoncyclicGeometric2020}
B.-J. Liu, S.-L. Su, M.-H. Yung, Nonadiabatic noncyclic geometric quantum
  computation in {{Rydberg}} atoms, Physical Review Research 2~(4) (2020)
  043130.
\newblock \href {https://doi.org/10.1103/PhysRevResearch.2.043130}
  {\path{doi:10.1103/PhysRevResearch.2.043130}}.

\bibitem{sunOnestepImplementationRydberg2022}
L.-N. Sun, F.-Q. Guo, Z.~Shan, M.~Feng, L.-L. Yan, S.-L. Su, One-step
  implementation of {{Rydberg}} nonadiabatic noncyclic geometric quantum
  computation in decoherence-free subspaces, Physical Review A 105~(6) (2022)
  062602.
\newblock \href {https://doi.org/10.1103/PhysRevA.105.062602}
  {\path{doi:10.1103/PhysRevA.105.062602}}.

\bibitem{tangFastEvolutionSingle2022}
G.~Tang, X.-Y. Yang, Y.~Yan, J.~Lu, Fast evolution of single qubit gate in
  non-adiabatic geometric quantum computing, PHYSICS LETTERS A 449~(128349)
  (Oct. 2022).
\newblock \href {https://doi.org/10.1016/j.physleta.2022.128349}
  {\path{doi:10.1016/j.physleta.2022.128349}}.

\bibitem{liangCompositeShortpathNonadiabatic2022}
Y.~Liang, P.~Shen, T.~Chen, Z.-Y. Xue, Composite short-path nonadiabatic
  holonomic quantum gates, PHYSICAL REVIEW APPLIED 17~(034015) (Mar. 2022).
\newblock \href {https://doi.org/10.1103/PhysRevApplied.17.034015}
  {\path{doi:10.1103/PhysRevApplied.17.034015}}.

\bibitem{liDynamicallyCorrectedNonadiabatic2021}
S.~Li, Z.-Y. Xue, Dynamically {{Corrected Nonadiabatic Holonomic Quantum
  Gates}}, Physical Review Applied 16~(4) (2021) 044005.
\newblock \href {https://doi.org/10.1103/PhysRevApplied.16.044005}
  {\path{doi:10.1103/PhysRevApplied.16.044005}}.

\bibitem{dridiOptimalRobustQuantum2020}
G.~Dridi, K.~Liu, S.~Gu{\'e}rin, Optimal {{Robust Quantum Control}} by
  {{Inverse Geometric Optimization}}, Physical Review Letters 125~(25) (2020)
  250403.
\newblock \href {https://doi.org/10.1103/PhysRevLett.125.250403}
  {\path{doi:10.1103/PhysRevLett.125.250403}}.

\bibitem{aiExperimentalRealizationNonadiabatic2020}
M.-Z. Ai, S.~Li, Z.~Hou, R.~He, Z.-H. Qian, Z.-Y. Xue, J.-M. Cui, Y.-F. Huang,
  C.-F. Li, G.-C. Guo, Experimental realization of nonadiabatic holonomic
  single-qubit quantum gates with optimal control in a trapped ion, PHYSICAL
  REVIEW APPLIED 14~(054062) (Nov. 2020).
\newblock \href {https://doi.org/10.1103/PhysRevApplied.14.054062}
  {\path{doi:10.1103/PhysRevApplied.14.054062}}.

\bibitem{xuExperimentalImplementationUniversal2020}
Y.~Xu, Z.~Hua, T.~Chen, X.~Pan, X.~Li, J.~Han, W.~Cai, Y.~Ma, H.~Wang, Y.~P.
  Song, Z.-Y. Xue, L.~Sun, Experimental implementation of universal
  nonadiabatic geometric quantum gates in a superconducting circuit, PHYSICAL
  REVIEW LETTERS 124~(230503) (Jun. 2020).
\newblock \href {https://doi.org/10.1103/PhysRevLett.124.230503}
  {\path{doi:10.1103/PhysRevLett.124.230503}}.

\bibitem{fengExperimentalRealizationNonadiabatic2013}
G.~Feng, G.~Xu, G.~Long, Experimental {{Realization}} of {{Nonadiabatic
  Holonomic Quantum Computation}}, PHYSICAL REVIEW LETTERS 110~(19) (May 2013).
\newblock \href {https://doi.org/10.1103/PhysRevLett.110.190501}
  {\path{doi:10.1103/PhysRevLett.110.190501}}.

\bibitem{abdumalikovExperimentalRealizationNonAbelian2013}
A.~A. Abdumalikov, Jr., J.~M. Fink, K.~Juliusson, M.~Pechal, S.~Berger,
  A.~Wallraff, S.~Filipp, Experimental realization of non-{{Abelian}}
  non-adiabatic geometric gates, NATURE 496~(7446) (2013) 482--485.
\newblock \href {https://doi.org/10.1038/nature12010}
  {\path{doi:10.1038/nature12010}}.

\bibitem{thomasRobustnessSinglequbitGeometric2011}
J.~T. Thomas, M.~Lababidi, M.~Tian, Robustness of single-qubit geometric gate
  against systematic error, PHYSICAL REVIEW A 84~(042335) (Oct. 2011).
\newblock \href {https://doi.org/10.1103/PhysRevA.84.042335}
  {\path{doi:10.1103/PhysRevA.84.042335}}.

\bibitem{zhengComparisonSensitivitySystematic2016}
S.-B. Zheng, C.-P. Yang, F.~Nori, Comparison of the sensitivity to systematic
  errors between nonadiabatic non-{{Abelian}} geometric gates and their
  dynamical counterparts, PHYSICAL REVIEW A 93~(032313) (Mar. 2016).
\newblock \href {https://doi.org/10.1103/PhysRevA.93.032313}
  {\path{doi:10.1103/PhysRevA.93.032313}}.

\bibitem{WOS:000824587200007}
M.-J. Liang, Z.-Y. Xue, Robust nonadiabatic geometric quantum computation by
  dynamical correction, PHYSICAL REVIEW A 106~(012603) (Jul. 2022).
\newblock \href {https://doi.org/10.1103/PhysRevA.106.012603}
  {\path{doi:10.1103/PhysRevA.106.012603}}.

\bibitem{dingDynamicalcorrectedNonadiabaticGeometric2023}
C.-Y. Ding, L.~Chen, L.-H. Zhang, Z.-Y. Xue, Dynamical-corrected nonadiabatic
  geometric quantum computation, FRONTIERS OF PHYSICS 18~(61304) (Dec. 2023).
\newblock \href {https://doi.org/10.1007/s11467-023-1322-2}
  {\path{doi:10.1007/s11467-023-1322-2}}.

\bibitem{liuSuperrobustNonadiabaticGeometric2021}
B.-J. Liu, Y.-S. Wang, M.-H. Yung, Super-robust nonadiabatic geometric quantum
  control, PHYSICAL REVIEW RESEARCH 3~(L032066) (Sep. 2021).
\newblock \href {https://doi.org/10.1103/PhysRevResearch.3.L032066}
  {\path{doi:10.1103/PhysRevResearch.3.L032066}}.

\bibitem{saffmanQuantumInformationRydberg2010}
M.~Saffman, T.~G. Walker, K.~Molmer, Quantum information with {{Rydberg}}
  atoms, REVIEWS OF MODERN PHYSICS 82~(3) (2010) 2313--2363.
\newblock \href {https://doi.org/10.1103/RevModPhys.82.2313}
  {\path{doi:10.1103/RevModPhys.82.2313}}.

\bibitem{saffmanQuantumComputingAtomic2016}
M.~Saffman, Quantum computing with atomic qubits and {{Rydberg}} interactions:
  Progress and challenges, JOURNAL OF PHYSICS B-ATOMIC MOLECULAR AND OPTICAL
  PHYSICS 49~(20) (Oct. 2016).
\newblock \href {https://doi.org/10.1088/0953-4075/49/20/202001}
  {\path{doi:10.1088/0953-4075/49/20/202001}}.

\bibitem{chenNonadiabaticGeometricQuantum2018}
T.~Chen, Z.-Y. Xue, Nonadiabatic {{Geometric Quantum Computation}} with
  {{Parametrically Tunable Coupling}}, PHYSICAL REVIEW APPLIED 10~(5) (Nov.
  2018).
\newblock \href {https://doi.org/10.1103/PhysRevApplied.10.054051}
  {\path{doi:10.1103/PhysRevApplied.10.054051}}.

\bibitem{WOS:000263313300001}
S.~Blanes, F.~Casas, J.~A. Oteo, J.~Ros, The {{Magnus}} expansion and some of
  its applications, PHYSICS REPORTS-REVIEW SECTION OF PHYSICS LETTERS 470~(5-6)
  (2009) 151--238.
\newblock \href {https://doi.org/10.1016/j.physrep.2008.11.001}
  {\path{doi:10.1016/j.physrep.2008.11.001}}.

\bibitem{WOS:000996294300003}
S.~L. Su, L.-N. Sun, B.~j.~Liu, L.~L. Yan, M.~H. Yung, W.~Li, M.~Feng, Rabi-
  and blockade-error-resilient all-geometric rydberg quantum gates, PHYSICAL
  REVIEW APPLIED 19~(044007) (Apr. 2023).
\newblock \href {https://doi.org/10.1103/PhysRevApplied.19.044007}
  {\path{doi:10.1103/PhysRevApplied.19.044007}}.

\bibitem{daemsRobustQuantumControl2013}
D.~Daems, A.~Ruschhaupt, D.~Sugny, S.~Guerin, Robust {{Quantum Control}} by a
  {{Single-Shot Shaped Pulse}}, PHYSICAL REVIEW LETTERS 111~(5) (Jul. 2013).
\newblock \href {https://doi.org/10.1103/PhysRevLett.111.050404}
  {\path{doi:10.1103/PhysRevLett.111.050404}}.

\bibitem{ruder2017overview}
S.~Ruder, An overview of gradient descent optimization algorithms (2017).
\newblock \href {http://arxiv.org/abs/1609.04747} {\path{arXiv:1609.04747}}.

\bibitem{WOS:000936540400002}
Y.~Liang, P.~Shen, L.-N. Ji, Z.-Y. Xue, State-independent nonadiabatic
  geometric quantum gates, PHYSICAL REVIEW APPLIED 19~(024051) (Feb. 2023).
\newblock \href {https://doi.org/10.1103/PhysRevApplied.19.024051}
  {\path{doi:10.1103/PhysRevApplied.19.024051}}.

\bibitem{liangStateIndependentNonadiabaticGeometric2023}
Y.~Liang, P.~Shen, L.-N. Ji, Z.-Y. Xue, State-{{Independent Nonadiabatic
  Geometric Quantum Gates}}, Phys. Rev. Appl. 19~(2) (2023) 024051.
\newblock \href {https://doi.org/10.1103/PhysRevApplied.19.024051}
  {\path{doi:10.1103/PhysRevApplied.19.024051}}.

\bibitem{lindbladGeneratorsQuantumDynamical1976}
G.~Lindblad, On the generators of quantum dynamical semigroups, Communications
  in Mathematical Physics 48~(2) (1976) 119--30.
\newblock \href {https://doi.org/10.1007/BF01608499}
  {\path{doi:10.1007/BF01608499}}.

\end{thebibliography}





\end{document}